%% file: report.tex
\documentclass[]{llncs}

\usepackage{pslatex}
\usepackage{amsmath}
\usepackage{amssymb}
\usepackage{leftidx}
\usepackage{epsfig}
\usepackage{paralist}
\usepackage{graphics}
\usepackage{stmaryrd}
\usepackage{txfonts}
\usepackage{framed}
\usepackage{makecell}
\usepackage{subfig}
\usepackage{wrapfig}
\usepackage[]{hyperref}
\usepackage{csquotes}
\usepackage{proof}

\usepackage[draft]{commenting}

\usepackage[version=0.96]{pgf}
\usepackage{tikz}
\usetikzlibrary{arrows,shapes,snakes,automata,backgrounds,petri,positioning}
\usepackage[latin1]{inputenc}

\declareauthor{ri}{Radu}{blue}
\declareauthor{mb}{Marius}{red}
\declareauthor{js}{Joseph}{magenta}
\declareauthor{cw}{Christoph}{teal}
\declareauthor{je}{Javier}{brown}

\input commands

\usepackage{thm-restate} 

\usepackage{setspace}

\newif\ifLongVersion\LongVersiontrue

\ifLongVersion
\else

\renewcommand{\proof}[1]{}
\fi

\begin{document}

\setlength{\belowdisplayskip}{2pt} \setlength{\belowdisplayshortskip}{1pt}
\setlength{\abovedisplayskip}{2pt} \setlength{\abovedisplayshortskip}{1pt}

\title{Verifying Safety Properties of Inductively Defined
  Parameterized Systems}

\author{Marius Bozga \and Radu Iosif}
\institute{Univ. Grenoble Alpes, CNRS, Grenoble INP\footnote{Institute of Engineering Univ. Grenoble Alpes}, Verimag}

\maketitle
\input abstract

\input body

\bibliographystyle{splncs03} \bibliography{refs}

\end{document}

%% file: commands.tex
\newtheorem{assumption}{Assumption}

\newcommand{\bigO}{\mathcal{O}}

\newcommand{\rbr}{{\bf ]\!]}}
\newcommand{\lbr}{{\bf [\![}}
\newcommand{\sem}[2]{\lbr #1 \rbr_{#2}}

\newcommand{\set}[1]{\left\{ #1 \right\}}
\newcommand{\tuple}[1]{\langle #1 \rangle}
\renewcommand{\vec}[1]{\mathbf #1}

\newcommand{\isdef}{\stackrel{\scriptscriptstyle{\mathsf{def}}}{=}}

\newcommand{\lang}[1]{\mathcal{L}({#1})}

\newcommand{\arrow}[2]{\xrightarrow{{\scriptscriptstyle #1}}_{{\scriptscriptstyle #2}}}

\newcommand{\nat}{{\bf \mathbb{N}}}

\newcommand{\proj}[2]{{#1}\!\!\downarrow_{{#2}}}

\renewcommand{\paragraph}[1]{\noindent{\bf #1}}

\newcommand{\vars}{\mathbb{V}_1}
\newcommand{\Vars}{\mathbb{V}_2}

\newcommand{\preds}{\mathsf{Pred}}

\newcommand{\fv}[1]{\mathrm{fv}({#1})}
\newcommand{\ins}[1]{\mathrm{inst}({#1})}
\newcommand{\height}[1]{\mathrm{ht}({#1})}

\newcommand{\pre}[1]{\leftidx{^\bullet}{\text{${#1}$}}}
\newcommand{\post}[1]{{#1}^\bullet}

\newcommand{\interv}[2]{[{#1},{#2}]}

\newcommand{\ids}{\mathbb{I}}

\newcommand{\idvar}{\xi}

\newcommand{\portset}{\mathbb{P}}
\newcommand{\stateset}{\mathbb{S}}
\newcommand{\abehtype}{\mathcal{B}}
\newcommand{\behtypeset}{\mathbb{B}}
\newcommand{\ports}{\mathcal{P}}
\newcommand{\states}{\mathcal{S}}
\newcommand{\errstates}{\mathsf{E}}
\newcommand{\initstate}{\mathcal{I}}
\newcommand{\rules}{\Delta}
\newcommand{\iports}{\mathsf{P}}
\newcommand{\istates}{\mathsf{S}}
\newcommand{\iinitstate}{\iota}
\newcommand{\irules}{\rightarrow}
\newcommand{\abeh}{\mathsf{B}}
\newcommand{\aint}{\pi}
\newcommand{\Inter}[1]{\mathsf{Inter}({#1})}
\newcommand{\aarch}{\mathcal{A}}
\newcommand{\archtype}{\Gamma}

\newcommand{\aterm}{\mathsf{t}}
\newcommand{\uterm}{\mathsf{u}}
\newcommand{\abehterm}{\mathsf{b}}
\newcommand{\apred}{\mathsf{A}}
\newcommand{\asys}{\mathcal{R}}
\newcommand{\arule}{\mathsf{r}}
\newcommand{\npred}[1]{\#_{\mathsf{pred}}({#1})}
\newcommand{\pred}[2]{\mathsf{pred}_{#1}({#2})}

\newcommand{\predset}{\mathbb{A}}
\newcommand{\instant}[2]{\left[{#1}\right]_{#2}}
\newcommand{\instances}[1]{\mathsf{inst}({#1})}
\newcommand{\canonic}[1]{{#1}^{\leadsto}}

\newcommand{\nodes}{\mathrm{nodes}}
\newcommand{\leaves}{\mathrm{leaves}}
\newcommand{\rtrees}[2]{\mathbb{T}_{#2}\big({#1}\big)}
\newcommand{\charterm}[2]{\mathfrak{C}_{#2}\left({#1}\right)}

\newcommand{\ainv}{\mathsf{I}}

\newcommand{\up}[1]{({#1},\uparrow)}
\newcommand{\dn}[1]{({#1},\downarrow)}

\newcommand{\wss}[1]{$\mathsf{WS}{#1}\mathsf{S}$}

\newcommand{\mso}{$\mathsf{MSO}$}

\newcommand{\wssroot}{\overline{\epsilon}}
\newcommand{\dom}{\mathrm{dom}}
\newcommand{\rng}{\mathrm{rng}}

\renewcommand{\succ}{\mathrm{succ}}

\newcommand{\size}[1]{\mathrm{size}({#1})}
\newcommand{\width}[1]{\mathrm{width}({#1})}


%% file: abstract.tex
We introduce a term algebra as a new formal specification language for
the coordinating architectures of distributed systems consisting of a
finite yet unbounded number of components. The language allows to
describe infinite sets of systems whose coordination between
components share the same pattern, using inductive definitions similar
to the ones used to describe algebraic data types or recursive data
structures. Further, we give a verification method for the parametric
systems described in this language, relying on the automatic synthesis
of structural invariants that enable proving general safety properties
(mutual exclusion, absence of deadlocks). The invariants are defined
using the \wss{\kappa}\ fragment of the monadic second order logic,
known to be decidable by a classical automata-logic connection. This
reduces the safety verification problem to checking satisfiability of
a \wss{\kappa}\ formula.

%% file: body.tex
\section{Introduction}

A fundamental principle in the design of a distributed system is the
separation between \emph{coordination} and \emph{behavior}
\cite{KramerMagee98}: the description of the coordinating architecture
of a software system states the components it is made of and how they
interact, whereas the components define the behavior they encapsulate
and specify which part of this behavior is visible in the
interface. The architecture then defines the interactions between the
interfaces of the components, ignoring the internal aspects of their
behavior.

Coordination is either \emph{endogenous}, i.e.\ making explicit use of
synchronization primitives in the code describing the behavior of the
components (e.g.\ semaphores, monitors, barriers, etc.) or
\emph{exogenous}, i.e.\ having global rules describing how the
components interact. A commonly perceived advantage of endogenous
coordination is that programmers do not have to explicitly build a
global coordination model. On the downside, endogenous coordination
does not cope well with formal aspects of concurrent/distributed
system design, for instance verification, because having a precise
description of the structure of interactions is typically needed in
order to automatically verify a parameterized system, in which the
number of replicated components is finite but the upper bound is not
known. More generally, exogenous coordination is a key enabler of the
study of coordination mechanisms and their properties, as attested by
the development of over a hundred architecture description languages
\cite{Bradbury04,Medvidovic97}.

Existing work on verification of parametric distributed systems
typically assumes hard-coded architectures, whose structure (but not
size) is fixed. For instance, the seminal work of German and Sistla
\cite{GermanSistla92} considers cliques, in which every component can
interact with every other component, whereas Emerson and Namjoshi
\cite{EmersonNamjoshi95} and Browne, Clarke and Grumberg
\cite{ClarkeGrumbergBrowne86} consider token-ring architectures, in
which each component interacts with its left and right neighbours
only. Most early results focus on the decidability and computational
complexity of verification problems such as safety (absence of error
configurations), depending tightly on the shape of the coordinating
architecture \cite{BloemJacobsKhalimovKonnovRubinVeithWidder15}.
Because decidability can only be obtained at the price of drastic
restrictions of the architectural pattern and of the communication
model (usually rendez-vous with a bounded number of participants),
more recent works go beyond the theoretical aspects and propose
practical semi-algorithmic methods, such as \emph{regular model
  checking}
\cite{KestenMalerMarcusPnueliShahar01,AbdullaHendaDelzannoRezine07} or
\emph{automata learning} \cite{ChenHongLinRummer17}. In such cases the
architectural pattern is implicitly determined by the class of
language recognizers: word automata encode pipelines or token-rings,
whereas tree automata are used to describe hierarchical
tree-structured architectures.

Among the first attempts at specifying architectures by logic is the
\emph{interaction logic} of Konnov et al. \cite{KonnovKWVBS16}, which
is a combination of Presburger arithmetic with monadic uninterpreted
function symbols (denoting communication ports), that can describe
cliques, stars and token-rings. They use first order logic without
successor functions, thus limiting the expressivity of the language
and excluding the possibility of describing more structured
architectures, such as pipelines, token-rings and tree-structured
hierachies. Such architectures can be described by an (undecidable)
second-order extension of the interaction logic
\cite{MavridouBBS17}. Our previous work on verifying safety properties
of architectures described using interaction logic(s) considers
interpreted successor functions that determine the shape of the
architecture: zero successors describe cliques
\cite{BozgaIosifSifakis19a}, one successor describe linear (pipeline,
token-ring) or star architectures (a single controller with many
slaves), whereas two or more successor functions describe tree-like
architectures \cite{DBLP:conf/tacas/BozgaEISW20}.

In this paper, we adhere to the exogenous coordination paradigm and
define a language for describing the architectures that coordinate the
interactions in a distributed system, parameterized
by\begin{inparaenum}[(i)]
\item the number of components of each type that are active in the
  system, e.g.\ a system with $n$ readers and $m$ writers, in which
  $n$ and $m$ are not known \`a~priori and 
\item the shape of the structure in which the interactions take place,
  e.g.\ a pipeline, ring, star, tree or, more general
  hypergraph-shaped structures.
\end{inparaenum}
We use a very simple syntax to describe the interactions between a
component and its immediate neighbours, together with a set of
inductive definitions that describe unbounded architectures, which
follow a common recursive pattern. The motivation behind using
inductive definitions is that recursive data structures, such as
algebraic datatypes \cite{DBLP:journals/jsat/BarrettST07} or memory
shapes \cite{DBLP:conf/lics/Reynolds02} are ubiquitous in programming,
hence programmers used to writing inductive specifications of data
structures could easily learn to write inductive specifications of
distributed component-based systems.

Specifying parameterized component-based systems by inductive
definitions is not new. \emph{Network grammars}
\cite{ShtadlerGrumberg89} use context-free grammar rules to describe
distributed systems with linear (pipeline, token-ring) architectures
obtained by composition of an unbounded number of concurrent
processes. Instead, we use predicate symbols of unrestricted arities
to describe architectural patterns that are, in general, more complex
than trees. Verification of network grammars against safety properties
requires the synthesis of \emph{network invariants}
\cite{WolperLovinfosse89}. Such network invariants can be computed by
rather costly fixpoint iterations \cite{LesensHalbwachsRaymond97} or
by abstracting the composition of a small bounded number of instances
\cite{KestenPnueliShaharZuck02}. Instead, our method uses lightweight
\emph{structural invariants}, that are shown to be easily inferred and
efficient in many practical examples
\cite{DBLP:conf/tacas/BozgaEISW20}.

For starters, let us consider the following specification of a system,
consising of components of type \emph{CType} with two interaction
ports, namely \emph{in} and \emph{out} and the behavior described by a
finite state machine with transitions $q_0 \arrow{\mathit{out}}{} q_1$
and $q_1 \arrow{\mathit{in}}{} q_0$. These components are arranged in
a ring, such that the \emph{out} port of a component is connected to
the \emph{in} port of its right neighbour, with the exception of the
last component, whose \emph{out} port connects to the \emph{in} port
of the first component (Fig. \ref{fig:ring}a). We specify this
architecture by means of a predicate $\mathit{Ring}()$ defined
inductively by the following rules:

\begin{figure}[t!]
  \caption{Recursive Specification of a Token-Ring System}
  \label{fig:ring}
  \vspace*{-.5\baselineskip}
  \centerline{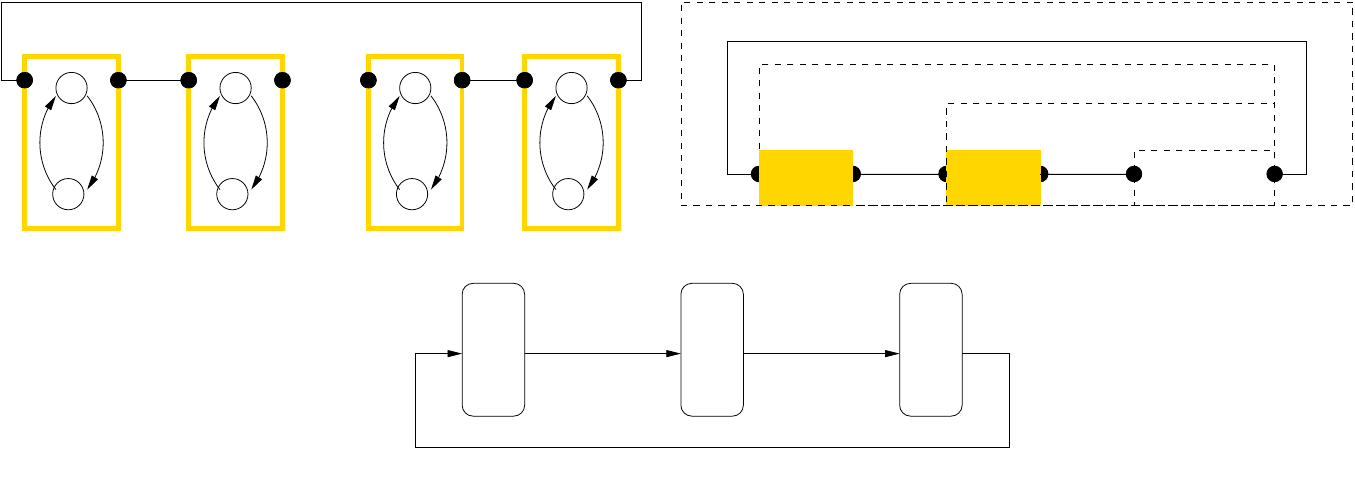}
\vspace*{-\baselineskip}
\end{figure}
\vspace*{-\baselineskip}
\begin{eqnarray}
\mathit{Ring}() & \leftarrow & 
\nu y_1 ~\nu y_2 ~.~ \tuple{\mathit{out}(y_2) \cdot \mathit{in}(y_1)} (\mathit{Chain}(y_1, y_2))
\label{rule:ring}
\end{eqnarray}
\begin{eqnarray}
\mathit{Chain}(x_1, x_2) & \leftarrow & 
\tuple{\mathit{out}(x_1) \cdot \mathit{in}(x_2)} (\mathit{CType}(x_1), \mathit{CType}(x_2)) 
\label{rule:chain1} \\
\mathit{Chain}(x_1, x_2) & \leftarrow & 
\nu y_1 ~.~ \tuple{\mathit{out}(x_1) \cdot \mathit{in}(y_1)} (\mathit{CType}(x_1), \mathit{Chain}(y_1, x_2))
\label{rule:chain2} 
\end{eqnarray}
Rule (\ref{rule:chain1}) says that the smallest chain consists of two
instances of type $\mathit{CType}$, namely $\mathit{CType}(x_1)$ and
$\mathit{CType}(x_2)$, such that the \emph{out} port of $x_1$ connects
to the \emph{in} port of $x_2$, described as $\mathit{out}(x_1) \cdot
\mathit{in}(x_2)$, where $x_1$ and $x_2$ are the formal arguments of the
rule. Rule (\ref{rule:chain2}) describes the inductive step, namely
that every chain stretching from $x_1$ to $x_2$ consists of a
component $\mathit{CType}(x_1)$ that interacts with a disjoint chain
from $y_1$ to $x_2$, where $y_1$ is an identifier different from every
other identifier in the system. Finally, rule (\ref{rule:ring}) closes
the ring by connecting the \emph{out} port of the last component $y_2$
to the \emph{in} port of the first component $y_1$, written as
$\mathit{out}(y_2) \cdot \mathit{in}(y_1)$. We refer to
Fig. \ref{fig:ring} for an illustration of the unfoldings of this set
of recursive definitions. Any system such as the one in
Fig. \ref{fig:ring}a is obtained by an application of rule
(\ref{rule:ring}), followed by $n$ applications of rule
(\ref{rule:chain2}), ending with an application of rule
(\ref{rule:chain1}). The first two applications of (\ref{rule:chain2})
following the application of (\ref{rule:ring}) are depicted in
Fig. \ref{fig:ring}b, with rule labels annotated in red.  Each
application of rule (\ref{rule:chain1}) creates a fresh variable,
denoted here as $y^1_1$, $y^2_1$, etc.

Having defined a language for specification of architectures, equipped
with a formal semantics that describes an architecture as an abstract
operator on finite-state behaviors, we move on to the \emph{parametric
  safety problem}, which is checking that the behavior of every
distributed system generated by an unfolding of a set of inductive
definitions stays clear of a set of unsafe configurations. For
instance, the behavior generated by the composition of three instances
of type $\mathit{CType}$ is depicted in Fig. \ref{fig:ring}c and the
safety property we check for is that in each state there is at least
one enabled transition.

Our method for proving safety relies on automatic invariant
synthesis. Like in our previous work
\cite{BozgaIosifSifakis19a,DBLP:conf/tacas/BozgaEISW20}, we use
structural invariants that can be derived directly from the behavioral
term and the recursive rewriting rules describing the system. The
verification method uses the invariant inference procedure to generate
a \wss{\kappa} formula that is unsatisfiable only if every system
described by the given inductive definitions is safe. Since
\wss{\kappa} is a decidable fragment of monadic second-order logic, we
use existing tools, such as \textsc{Mona} \cite{Mona} for proving
(parametric) safety. We have implemented the invariant synthesis in a
prototype tool and experimented our method on a number of parametric
component-based systems with non-trivial architectural patterns, such
as trees with root links, trees with linked leaves, token-rings
with(out) a main controller (star), etc. For space reasons, the proofs
of the technical results are given in \cite{BozgaIosif20Arxiv}.

\section{Behaviors and Architectures}
\label{sec:prelim}

This section introduces the preliminary definitions of a
(finite-state) behavior and a bounded architecture, before defining
behavioral types, that are the first ingredient of a formal definition
of parametric component-based systems. Given sets $A$ and $B$, we
denote by $A \mapsto B$ the set of total functions from $A$ into
$B$. Partial mappings from $A$ to $B$ are denoted as $f : A
\rightharpoonup B$, where $\dom(f) \isdef \set{a \in A \mid f(a)
  \text{ is defined}}$ is the domain and $\rng(f) \isdef \set{f(a)
  \mid a \in \dom(f)}$ is the range of $f$.

Let $\portset = \set{a, b, \ldots}$ and $\stateset = \set{s, t,
  \ldots}$ be countably infinite sets of \emph{ports} and
\emph{states}, respectively. A \emph{configuration} $\sigma \subseteq
\stateset$ is a finite set of states. A \emph{behavior} is a tuple
$\abeh = \tuple{\iports, \istates, \iinitstate, \irules}$, where
$\iports \subseteq \portset$ and $\istates \subseteq \stateset$ are
finite sets of ports and states, respectively, $\iinitstate \subseteq
\istates$ denotes the initial configuration and $\irules \subseteq
2^\istates \times 2^\iports \times 2^\istates$ is a set of
\emph{transitions} denoted as $\sigma \arrow{\aint}{} \tau$, for some
configurations $\sigma,\tau \subseteq \istates$ and some set of ports
$\aint \subseteq \iports$. We assume the existence of an \emph{idling}
transition $\sigma \arrow{\emptyset}{} \sigma$, for each configuration
$\sigma \subseteq \istates$ and denote by $\iports_\abeh$,
$\istates_\abeh$, $\iinitstate_\abeh$ and $\irules_\abeh$ the ports,
states, initial configuration and transitions of $\abeh$,
respectively. An \emph{execution path} of $\abeh$ is a sequence of
transitions $\sigma_1 \arrow{\aint_1}{\abeh} \sigma_2
\arrow{\aint_2}{\abeh} \ldots$ A configuration $\sigma \subseteq
\istates$ is \emph{reachable} in $\abeh$ iff $\abeh$ has a finite
execution path starting with $\iota$ and leading to $\sigma$; $\abeh$
is \emph{safe} w.r.t. a set of configurations $\errstates$ iff no
configuration from $\errstates$ is reachable in $\abeh$.

Given two behaviors $\abeh_i = \tuple{\iports_i, \istates_i,
  \iinitstate_i, \irules_i}$, for $i=1,2$, such that $\istates_1 \cap
\istates_2 = \emptyset$ and $\iports_1 \cap \iports_2 = \emptyset$, we
define their \emph{product} as $\abeh_1 \parallel \abeh_2 \isdef
\tuple{\iports_1 \cup \iports_2, \istates_1 \cup \istates_2,
  \iinitstate_1 \cup \iinitstate_2, \irules_{\abeh_1 \parallel
    \abeh_2}}$, where $\irules_{\abeh_1 \parallel \abeh_2}$ is the
smallest set of transitions defined by the rule
(\ref{infrule:prod}). Intuitively, the product of two behaviors
consists of any transition that belongs to either one of the two
behaviors or a combined transition using the ports of both transitions
in a joint action\footnote{In particular, each transition $\sigma_1
  \arrow{\aint_1}{1} \tau_1$ induces a transition $(\sigma_1 \cup
  \sigma_2) \arrow{\aint_1}{} (\tau_1 \cup \sigma_2)$ due to the
  idling transition $\sigma_2 \arrow{\emptyset}{} \sigma_2$.}. Since
$\parallel$ is commutative and associative, we write $\abeh_1
\parallel \ldots \parallel \abeh_n$ instead of $(\abeh_1 \parallel
\abeh_2) \parallel \ldots \parallel \abeh_n$. 

\begin{center}
  \begin{minipage}{.5\textwidth}
    \begin{equation}\label{infrule:prod}
      \infer[]{
        (\sigma_1 \cup \sigma_2) \arrow{\aint_1 \cup~ \aint_2}{\abeh_1 \parallel \abeh_2} (\tau_1 \cup \tau_2)
      }{
        \sigma_i \arrow{\aint_i}{i} \tau_i,~ i = 1,2
      }
    \end{equation}    
  \end{minipage}
  \begin{minipage}{.49\textwidth}
    \begin{equation}\label{infrule:arch}
      \infer[]{
        \sigma \arrow{\aint}{\gamma(\abeh_1, \ldots, \abeh_n)} \tau
      }{
        \sigma \arrow{\aint}{\abeh_1 \parallel \ldots \parallel \abeh_n} \tau,~ \aint \in \gamma
      }
    \end{equation}
  \end{minipage}
\end{center}

The product of behaviors (\ref{infrule:prod}) is, in general, too
permissive and allows unsafe executions. We refine this operator to
achieve a desired level of safety, by means of \emph{architectures}, a
central notion in the rest of this paper, defined below:
\begin{definition}\label{def:arch}
An \emph{interaction} $\aint \subseteq \portset$ is a finite set of
ports. An \emph{architecture} $\gamma \subseteq 2^\portset$ is a
finite set of interactions.
\end{definition}
Just as the product of behaviors (\ref{infrule:prod}), an architecture
can be viewed as a commutative and associative operator, whose
application to the set of behaviors $\set{\abeh_i = \tuple{\iports_i,
    \istates_i, \iinitstate_i, \irules_i}}_{i=1}^n$ is the behavior
\(\gamma(\abeh_1, \ldots, \abeh_n) \isdef
\langle\bigcup_{i=1}\iports_i, \bigcup_{i=1}^n\istates_n,
\bigcup_{i=1}^n\iinitstate_i, \irules_{\gamma(\abeh_1, \ldots,
  \abeh_n)}\rangle\), where $\irules_{\gamma(\abeh_1, \ldots,
  \abeh_n)}$ is the least set of transitions defined by the rule
(\ref{infrule:arch}). The architecture $\gamma$ simply restricts the
transitions of the product $\abeh_1 \parallel \ldots \parallel
\abeh_n$ to the ones labeled with an interaction from $\gamma$. Note
that the arity of $\gamma$ is not fixed, i.e.\ $\gamma(\abeh_1,
\ldots, \abeh_n)$ is defined, for all $n\geq1$.

In the rest of this paper, we are concerned with systems consisting of
an unbounded number of replicated behaviors, that belong to a fairly
small number of patterns, called component types. Let $\ids = \set{i,
  j, \ldots}$ be a countably infinite set of \emph{identifiers}. A
\emph{component type} is a tuple $\abehtype = \tuple{\ports_\abehtype,
  \states_\abehtype, \initstate_\abehtype, \rules_\abehtype}$, where
$\ports_\abehtype \subseteq \ids \mapsto \portset$ and
$\states_\abehtype \subseteq \ids \mapsto \stateset$ are finite sets
of total functions mapping identifiers to ports and states,
respectively, $\initstate_\abehtype \in \states$ denotes initial
states and $\rules_\abehtype \subseteq (\ids \mapsto \states) \times
(\ids \mapsto \portset) \times (\ids \mapsto \states)$ is a finite set
of \emph{transition rules} of the form $S \arrow{P}{} T$. In addition,
we require that, for any $P, Q \in \ports$ [$S,T \in \states$] and $i,
j \in \ids$, such that $P(i)=Q(j)$ [$S(i)=T(j)$], we have $P=Q$
[$S=T$] and $i=j$, i.e.\ all elements of $\ports_\abehtype$
[$\states_\abehtype$] are injective functions with pairwise disjoint
ranges.

Given a component type $\abehtype = \tuple{\ports, \states,
  \initstate, \rules}$ and an identifier $i \in \ids$, the behavior
$\abehtype(i) \isdef \tuple{\set{P(i) \mid P \in \ports}, \set{S(i)
    \mid S \in \states}, \set{\initstate(i)}, \{\set{S(i)}
  \arrow{\set{P(i)}}{} \set{T(i)} \mid S \arrow{P}{} T \in \rules\}}$
is called the $i$-th \emph{instance} of $\abehtype$. \ifLongVersion As
one would expect, each reachable configuration of an instance consists
of one state and each transition of an instance is labeled with a
singleton set of ports. \fi Note that $\iports_{\abehtype(i)} \cap
\iports_{\abehtype(j)} = \emptyset$ and $\istates_{\abehtype(i)} \cap
\istates_{\abehtype(j)} = \emptyset$, for any $i \neq j \in \ids$. 

In the rest of this paper, we consider a fixed set $\behtypeset$ of
component types, such that $\ports_{\abehtype_1} \cap
\ports_{\abehtype_2} = \emptyset$ and $\states_{\abehtype_1} \cap
\states_{\abehtype_2} = \emptyset$, for any $\abehtype_1, \abehtype_2
\in \behtypeset$.

\ifLongVersion
\begin{figure}[htb]
  \centerline{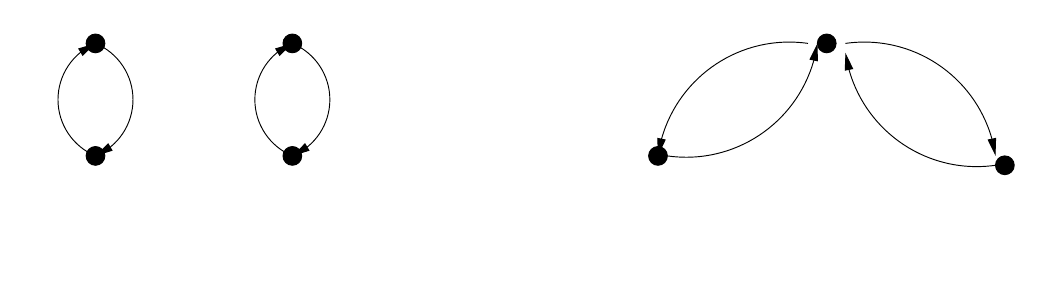}
  \vspace*{-.5\baselineskip}
  \caption{Component Types $\mathit{Task}$ and $\mathit{Lock}$ (a).
    Semantics of the Composition $\{\{\mathit{acq}(i), \mathit{lock}(k)\},
    \{\mathit{acq}(j),\mathit{lock}(k)\},
    \{\mathit{rel}(i),\mathit{unlock}(k)\}, \{\mathit{rel}(j),
    \mathit{unlock}(k)\}\}(\mathit{Task}(i), \mathit{Task}(j)$,
    $\mathit{Lock}(k))$ (b)}
  \label{fig:behavioral-types}
  \vspace*{-\baselineskip}
\end{figure}

\begin{example}\label{ex:behavioral-types}
Figure \ref{fig:behavioral-types}a depicts two component types
$\mathit{Task}$ and $\mathit{Lock}$, whereas Figure
\ref{fig:behavioral-types}b shows the composition of three instances
$\mathit{Task}(i)$, $\mathit{Task}(j)$ and $\mathit{Lock}(k)$, via the
architecture $\set{\set{\mathit{acq}(i), \mathit{lock}(k)},
  \set{\mathit{acq}(j),\mathit{lock}(k)},
  \set{\mathit{rel}(i),\mathit{unlock}(k)}, \set{\mathit{rel}(j),
    \mathit{unlock}(k)}}$ and $i, j, k \in \ids$ are pairwise distinct
identifiers. \hfill$\blacksquare$
\end{example}
\fi

\section{A Term Algebra of Behaviors}
\label{sec:term-algebra}

In this section we introduce a recursive term algebra for describing
the behaviors resulting from the composition of an unbounded number of
component type instances. Let $\vars$ be a countably infinite set of
first-order variables and $\predset$ be a countably infinite set of
\emph{predicates}, where $\#(\apred) \geq 0$ denotes the arity of
$\apred \in \predset$. The following syntax generates \emph{behavioral
  terms} inductively, starting with the $\abehterm$ non-terminal:
\[\begin{array}{ll}
P \in \ports,~ x \in \vars,~ 
i \in \ids,~ \abehtype \in \behtypeset,~ \apred \in \predset \\
\idvar ::= x \mid i 
\hspace*{1cm}
\archtype ::= P(\idvar) \mid \archtype_1 \cdot \archtype_2 \mid \archtype_1 + \archtype_2 & \text{ architecture specifications} \\
\abehterm ::= \abehtype(\idvar) \mid \tuple{\archtype}(\abehterm_1, \ldots, \abehterm_n) 
\mid \nu x ~.~ \abehterm_1 \mid \apred(\idvar_1, \ldots, \idvar_{\#(\apred)}) & \text{ behavioral terms}
\end{array}\]
A variable $x$ occurring in a behavioral term $\abehterm$ is said to
be \emph{free} if it does not occur in the scope of some subterm of
the form $\nu x ~.~ \abehterm_1$ and \emph{bound} otherwise. In the
following, we assume that all bound variables occurring in a term are
pairwise distinct and distinct from the free variables. Note that this
assumption loses no generality because terms obtained by
$\alpha$-conversion (renaming of bound variables) are assumed to be
equivalent. A term $\abehterm$ is said to be \emph{closed} if
$\fv{\abehterm} = \emptyset$, \emph{predicate-less} if no predicates
from $\predset$ occur in $\abehterm$ and \emph{ground} if no variable,
either free or bound, occurs in $\abehterm$. A term
$\abehtype(\idvar)$ is called an \emph{instance atom} and a term
$\apred(\idvar_1, \ldots, \idvar_n)$ is called a \emph{predicate
  atom}. We denote by $\instances{\abehterm}$ the set of instance
atoms of $\abehterm$, by $\npred{\abehterm}$ the number of occurrences
of predicate atoms and by $\pred{j}{\abehterm}$, $j \in
\interv{0}{\npred{\abehterm}-1}$, the predicate atom that occurs
$j$-th in $\abehterm$, in some predefined order of the syntax tree
nodes of $\abehterm$. \ifLongVersion We write $\size{\abehterm}$ for
the number of occurrences of symbols in $\abehterm$. \fi

\begin{figure}[t!]
\caption{Tree Architecture with Leaves Linked in a Token-Ring}
\label{fig:tll}
\vspace*{-.5\baselineskip}
\centerline{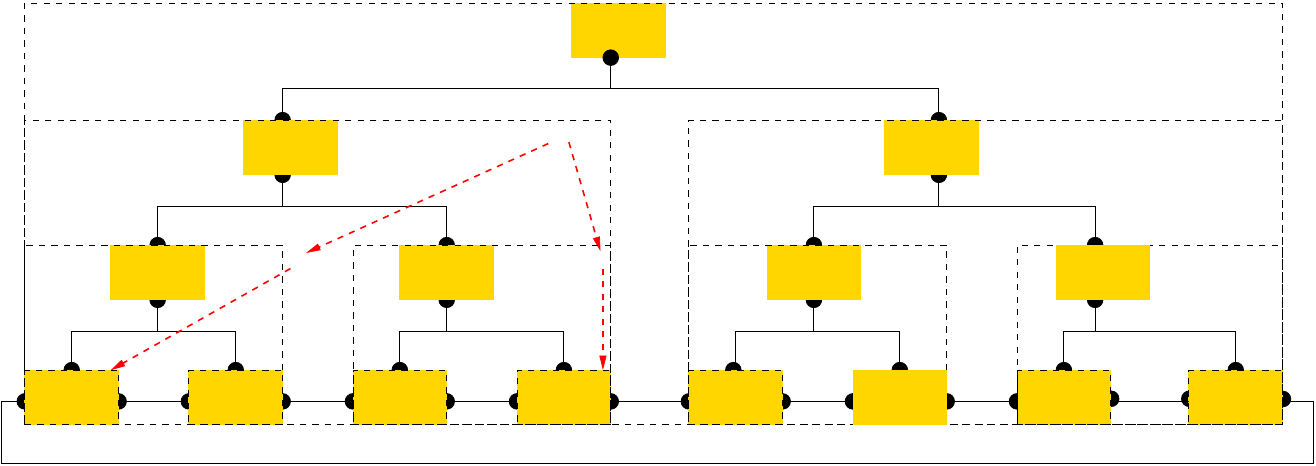}
\vspace*{-\baselineskip}
\end{figure}

A symbol $\idvar \in \vars\cup\ids$ is \emph{instantiated} in a
behavioral term $\abehterm$ if $\abehtype(\idvar)$ is a subterm of
$\abehterm$, for some component type $\abehtype$, and we denote by
$\ins{\abehterm}$ the set of symbols instantiated in $\abehterm$. Note
that a symbol (variable or identifier) may occur in a term without
being instantiated. For example, both identifiers $i$ and $j$ occur
within the term $\tuple{\mathit{out}(i) \cdot
  \mathit{in}(j)}(\mathit{CType}(j))$, but only $j$ is instantiated by
the atom $\mathit{CType}(j)$. A behavioral term $\abehterm$ is
\emph{well-instantiated} if every identifier occurring in $\abehterm$
is instantiated at most once. For example, the following term is not
well-instantiated, because $i$ is instantiated twice in
$\tuple{\mathit{out}(i) \cdot \mathit{in}(j)}(\mathit{CType}(i),
\tuple{\mathit{in}(i) \cdot \mathit{out}(j)}(\mathit{CType}(j),
\mathit{CType}(i)))$.

A \emph{substitution} is a partial function $\eta : \vars
\rightharpoonup \vars \cup \ids $ mapping variables into either
variables or identifiers. A substitution $\eta$ is \emph{ground} if
$\rng(\eta) \subseteq \ids$. We denote by $[\idvar_1/x_1, \ldots,
  \idvar_n/x_n]$ the substitution mapping each $x_i \in \vars$ into
$\idvar_i \in \vars \cup \ids$, for all $i \in \interv{1}{n}$, and
undefined everywhere else. The application of a substitution $\eta$ to
a behavioral term $\abehterm$ is the term $\abehterm\eta$ in which
every free occurrence of a variable $x \in \fv{\abehterm} \cap
\dom(\eta)$ has been replaced by $\eta(x)$. Note that substitutions
only apply to the free variables of the term.

Given a predicate-less behavioral term $\abehterm$ and a ground
substitution $\eta$, such that $\fv{\abehterm} \subseteq \dom(\eta)$,
the \emph{ground set} of $\abehterm$ is the set
$\instant{\abehterm}{\eta}$ of ground terms, defined inductively:
\[\begin{array}{rcl}
\instant{\abehtype(x)}{\eta} & \isdef & \set{\abehtype(\eta(x))}
\hspace*{5mm}
\instant{\abehtype(i)}{\eta} \isdef \set{\abehtype(i)}
\hspace{5mm}
\instant{\nu x ~.~ \abehterm_1}{\eta} \isdef 
\bigcup_{i \in \ids \setminus \rng(\eta)} \instant{\abehterm_1}{\eta[x \leftarrow i]} \\
\instant{\tuple{\archtype}(\abehterm_1, \ldots, \abehterm_n)}{\eta} & \isdef & 
\bigcup\{\tuple{\archtype}(\aterm_1, \ldots, \aterm_n)\eta \mid \forall 1 \leq k < \ell \leq n ~.~ 
\aterm_k \in \instant{\abehterm_k}{\eta} \wedge \ins{\aterm_k} \cap \ins{\aterm_\ell} = \emptyset\} 
\end{array}\]
Whenever $\abehterm$ is closed, $\eta$ can be omitted and its ground
set can be written $\instant{\abehterm}{}$. Note that the definition
of the ground set prevents multiple instantiation of the same
identifier.
\ifLongVersion 
For instance, in Fig. \ref{fig:behavioral-types}b, the
ground term $\mathit{acq}(i) \cdot \mathit{lock}(k) + \mathit{acq}(j)
\cdot \mathit{lock}(k) + \mathit{rel}(i) \cdot \mathit{unlock}(k) +
\mathit{rel}(j) \cdot \mathit{unlock}(k)(\mathit{Task}(i),
\mathit{Task}(j)$, $\mathit{Lock}(k))$ belongs to the ground set of
the behavioral term $\nu x \nu y \nu z ~.~ \mathit{acq}(x) \cdot
\mathit{lock}(z) + \mathit{acq}(y) \cdot \mathit{lock}(z) +
\mathit{rel}(x) \cdot \mathit{unlock}(z) + \mathit{rel}(y) \cdot
\mathit{unlock}(z)(\mathit{Task}(x), \mathit{Task}(y)$,
$\mathit{Lock}(z))$. 
\fi

The meaning of a ground architecture specification $\archtype$, built
from ports $P(i)$, $P \in \ports$, $i \in \ids$, using the
constructors $+$ and $\cdot$, is the architecture $\sem{\archtype}{}
\subseteq 2^{\ports}$, defined inductively:
\[\begin{array}{c}
\sem{P(i)}{} \isdef \set{\set{P(i)}} 
\hspace*{2mm} 
\sem{\archtype_1 + \archtype_2}{} \isdef \sem{\archtype_1}{} \cup \sem{\archtype_2}{} 
\hspace*{2mm}
\sem{\archtype_1 \cdot \archtype_2}{} \isdef \set{I_1 \cup I_2 \mid I_i \in \sem{\archtype_i}{},~ i = 1,2}
\end{array}\]
Note that the $+$ and $\cdot$ constructors are both commutative and
associative. Moreover, $\cdot$ distributes over $+$, thus each ground
architecture specification can be equivalently written $\archtype =
\sum_{k=1}^m \prod_{\ell=1}^{r_k} P_{k\ell}(i_{k\ell})$, where
$P_{k\ell} \in \ports$ and $i_{k\ell} \in \ids$, for all $k \in
\interv{1}{m}$ and $\ell \in \interv{1}{r_k}$.

We extend ground sets from predicate-less terms to terms with
predicate occurrences, by recursively replacing predicate subterms by
terms given by a set of rewriting rules (called a \emph{rewriting
  system}) of the form $\apred(x_1, \ldots, x_{\#(\apred)}) \leftarrow
\abehterm$, where $\abehterm$ is a behavioral term, such that
$\fv{\abehterm} \subseteq \set{x_1, \ldots, x_{\#(\apred)}}$. For
conciseness, we write $\apred(x_1, \ldots, x_{\#(\apred)})
\leftarrow_\asys \abehterm$ instead of $\apred(x_1, \ldots,
x_{\#(\apred)}) \leftarrow \abehterm \in \asys$. \ifLongVersion The
\emph{size} of $\asys$ is $\size{\asys} \isdef \sum_{\apred(x_1,
  \ldots, x_{\#(\apred)}) \leftarrow_\asys \abehterm}
\size{\abehterm}$ and its \emph{width} is $\width{\asys} \isdef
\max_{\apred(x_1, \ldots, x_{\#(\apred)}) \leftarrow_\asys
  \abehterm}\size{\abehterm}$. \fi

\begin{example}\label{ex:tll}
The following example describes, by the term $\mathit{Root}()$, a tree
architecture in which parents communicate with their children and, in
addition, all nodes on the frontier communicate via a token-ring. The
inner nodes in the tree have component type $\mathit{NType}$, with
associated ports $\mathit{req}$ and $\mathit{reply}$, whereas the
leaves have type $\mathit{LType}$, with associated ports
$\mathit{reply}$, $\mathit{in}$ and $\mathit{out}$.
\begin{eqnarray}
\mathit{Root}() & \leftarrow & \nu r~ \nu n_1 ~\nu l_1 ~\nu r_1 ~\nu n_2 ~\nu l_2 ~\nu r_2 ~.~ \nonumber \\
&& \tuple{\mathit{req}(r) \cdot \mathit{reply}(n_1) \cdot \mathit{reply}(n_2) + 
\mathit{out}(r_1) \cdot \mathit{in}(l_2) + \mathit{out}(r_2) \cdot \mathit{in}(l_1)} \nonumber \\
&& (\mathit{Ntype}(r), \mathit{Node}(n_1, l_1, r_1), \mathit{Node}(n_2, l_2, r_2)) \label{rule:tll-root} \\
\mathit{Node}(n, l, r) & \leftarrow & \nu n_1 ~\nu r_1 ~\nu n_2 ~\nu l_2 ~.~ \nonumber 
\tuple{\mathit{req}(n) \cdot \mathit{reply}(n_1) \cdot \mathit{reply}(n_2) + \mathit{out}(r_1) \cdot \mathit{in}(l_2)} \nonumber \\
&& (\mathit{NType}(n), \mathit{Node}(n_1, l, r_1), \mathit{Node}(n_2, l_2, r)) \label{rule:tll1} \\
\mathit{Node}(n, l, r) & \leftarrow & 
\tuple{\mathit{req}(n) \cdot \mathit{reply}(l) \cdot \mathit{reply}(r) + \mathit{out}(l) \cdot \mathit{in}(r)} \nonumber \\
&& (\mathit{NType}(n), \mathit{Leaf}(l), \mathit{Leaf}(r)) \label{rule:tll4} \\
\mathit{Leaf}(n) & \leftarrow & \mathit{LType}(n) \label{rule:tll5}
\end{eqnarray}
We refer to Fig. \ref{fig:tll} for a depiction of the unfolding of the
above rewriting rules and of the resulting architecture. The labels of
the rewriting rules applied at each rewriting step are marked in
red. For readability, we superscript each bound variable introduced by
a rule with the node of the rewriting tree where this rule was
applied.
Each rule (\ref{rule:tll-root}-\ref{rule:tll4}) creates an interaction
between the parent node ($\mathit{req}$) and its children
($\mathit{reply}$) and the leaf rules (\ref{rule:tll5}) also creates
interactions between siblings of the form
$\{\mathit{out}(i),\mathit{in}(j)\}$. In addition, the initial rule
(\ref{rule:tll-root}) closes the ring of leaves, via the interactions
$\{\mathit{out}(r^\epsilon_2), \mathit{in}(l^\epsilon_1)\}$ and
$\{\mathit{out}(r^\epsilon_1), \mathit{in}(l^\epsilon_2)\}$, where the
parameters $l^\epsilon_1$ and $r^\epsilon_2$ are instantiated in the
left- and right-most leaves and $r^\epsilon_1$ ($l^\epsilon_2$) in the
right-most (left-most) leaf of the left (right) subtree.
\hfill$\blacksquare$
\end{example}

For technical convenience, we place the steps of an rewriting sequence
in a tree, whose nodes are labeled by rewriting rules. Formally, a
{\em tree} $\mathcal{T}$ is defined by a set $\nodes(\mathcal{T})$ and
a function mapping each node $w\in \nodes(\mathcal{T})$ to its {\em
  label}, denoted by $\mathcal{T}(w)$. The set $\nodes(\mathcal{T})$
is a finite subset of $\nat^*$, where $\nat^*$ is the set of finite
sequences of non-negative integers, such that $wi \in
\nodes(\mathcal{T})$ for some $i \in \nat \setminus \set{0}$ only if
$w \in \nodes(\mathcal{T})$ and $wj \in \nodes(\mathcal{T})$ for all
$j\in\interv{0}{i-1}$. The \emph{root} of $\mathcal{T}$ is the empty
sequence $\epsilon$, the {\em children} of a node $w\in
\nodes(\mathcal{T})$ are the nodes $wi \in \nodes(\mathcal{T})$, where
$i \in \nat$, and the {\em parent} of a node $wi$ with $i \in \nat$ is
$w$ (the root $\epsilon$ has no parent). The \emph{leaves} of $\mathcal{T}$ are
the nodes in $\leaves(\mathcal{T}) \isdef \set{w \in
  \nodes(\mathcal{T}) \mid w.0 \not\in \nodes(\mathcal{T})}$. The
\emph{subtree} of $\mathcal{T}$ rooted at $w$ is defined as
$\proj{\mathcal{T}}{w}(w') \isdef \mathcal{T}(ww')$, for all $w' \in
\nodes(\proj{\mathcal{T}}{w}) \isdef \{ w' \mid ww' \in
\nodes(\mathcal{T}) \}$.

\begin{definition}\label{def:rewriting-tree}
  Given a rewriting system $\asys$ and a closed behavioral term
  $\abehterm$, a \emph{rewriting tree} for $\abehterm$ is a tree
  $\mathcal{T}$ such that $\mathcal{T}(\epsilon) =
  \left(\apred_\abehterm() \leftarrow \abehterm\right)$, where
  $\apred_\abehterm$ is a predicate symbol of zero arity, that does
  not occur in $\asys$ and, for all $w \in \dom(\mathcal{T})$, such
  that $\mathcal{T}(w) = \left(\apred_w(x_1, \ldots, x_{\#(\apred_w)})
  \leftarrow_\asys \abehterm_w\right)$: \begin{compactenum}
  \item\label{it1:rewriting-tree} for all $i \in
    \interv{0}{\npred{\abehterm_w}-1}$, if $\pred{i}{\abehterm_w} =
    \apred_{wi}(y_1, \ldots, y_{\#(\apred_{wi})})$ then $wi \in
    \nodes(\mathcal{T})$ and $\mathcal{T}(wi) = \apred_{wi}(x_1,
    \ldots, x_{\#(\apred_{wi})}) \leftarrow_\asys \abehterm_{wi}$, for
    some behavioral term $\abehterm_{wi}$,
  \item\label{it2:rewriting-tree} for all $i \geq
    \npred{\abehterm_w}$, we have $wi \not\in \nodes(\mathcal{T})$.
  \end{compactenum}
  We denote $\asys_\abehterm \isdef \asys \cup \set{\apred_\abehterm()
    \leftarrow \abehterm}$ and by $\rtrees{\abehterm}{\asys}$ the set
  of rewriting trees for $\abehterm$ in $\asys$.
\end{definition}
Note the addition of a fresh rule $\apred_\abehterm() \leftarrow
\abehterm$ to $\asys$, that is required for a uniform labeling of the
tree with rules. \ifLongVersion Moreover, since $\abehterm$ is assumed
to be closed, the condition $\#(\apred_\abehterm) = 0$ is consistent
with the definition of a rewriting rule, because $\fv{\abehterm}$ is
always required to be contained in the argument list of
$\apred_\abehterm$. \fi For instance, Fig. \ref{fig:tll} shows a
balanced binary rewriting tree, whose root is labeled by rule
(\ref{rule:tll-root}), second and third level nodes are labeled by
rules (\ref{rule:tll1}) and (\ref{rule:tll4}) respectively, and leaves
are labeled by rule (\ref{rule:tll5}). A rewriting tree $\mathcal{T}
\in \rtrees{\abehterm}{\asys}$ corresponds to a predicate-less
\emph{characteristic term} $\charterm{\mathcal{T}}{}$ defined
inductively on the structure of $\mathcal{T}$:
\begin{definition}\label{def:char-term}
Given a behavioral term $\abehterm$ and a rewriting tree $\mathcal{T}
\in \rtrees{\abehterm}{\asys}$, for each $w \in \nodes(\mathcal{T})$,
such that $\mathcal{T}(w) = \left(\apred_w(x_1, \ldots,
x_{\#(\apred_w)}) \leftarrow_\asys \abehterm_w\right)$, we define
$\charterm{\proj{\mathcal{T}}{w}}{}$ as the term obtained by replacing
each predicate atom $\pred{i}{\abehterm} = \apred_{wi}(y_1, \ldots,
y_{\#(\apred_{wi})})$ by the term
$\charterm{\proj{\mathcal{T}}{wi}}{}[y_1/x_1, \ldots,
  y_{\#(\apred_i)}/x_{\#(\apred_i)}]$, for all $i \in
\interv{0}{\npred{\abehterm_w}-1}$. We define
$\charterm{\mathcal{T}}{} \isdef
\charterm{\proj{\mathcal{T}}{\epsilon}}{}$ and
$\charterm{\abehterm}{\asys} \isdef \set{\charterm{\mathcal{T}}{} \mid
  \mathcal{T} \in \rtrees{\abehterm}{\asys}}$ the set of
predicate-less terms generated by $\asys$ from $\abehterm$.
\end{definition}
Intuitively, the characteristic term of a rewriting tree is the
predicate-less term obtained by replacing all predicate instances by
the bodies of their corresponding rewriting rules from the tree.  We
define the ground set of behavioral term (possibly with predicate
atoms) as \(\instant{\abehterm}{\eta,\asys} \isdef \bigcup_{\aterm \in
  \charterm{\abehterm}{\asys}} \instant{\aterm}{\eta}\) and avoid
mentioning $\eta$ when $\abehterm$ is closed.

Next, we define a semantic operator $\sem{.}{}$ that maps ground
behavioral terms to behaviors. The idea is that nested terms, such as
e.g.\ $\tuple{\mathit{out}(x) \cdot \mathit{in}(y)}(\mathit{CType}(x),
\tuple{\mathit{in}(x) \cdot \mathit{out}(y)}(\mathit{CType}(y)))$ are
not be dealt with compositionally; instead, all the (partial)
architectures that occur within subterms of a behavioral term are
first joined into a top-level architecture that applies, at the same
time, to all instances in the term. Formally, we define the following
\emph{flattening} relation on behavioral terms:
\begin{equation}\label{eq:arch-sum}
\tuple{\archtype_1}(\tuple{\archtype_2}(\abehterm_1, \ldots \abehterm_i),
\abehterm_{i+1}, \ldots \abehterm_n) \leadsto \tuple{\archtype_1 +
\archtype_2}(\abehterm_1, \ldots, \abehterm_n)
\end{equation}
Note that the order of the arguments $\tuple{\archtype_2}(\abehterm_1,
\ldots \abehterm_i), \abehterm_{i+1}, \ldots, \abehterm_n$ of
$\archtype_1$ is not important. It is easy to see that every chain
$\aterm_1 \leadsto \aterm_2 \leadsto \ldots$ is finite, because
\ifLongVersion $\height{\aterm_1} > \height{\aterm_2} > \ldots > 0$,
where $\height{\aterm} > 0$ is defined inductively on the structure of
$\aterm$ as $\height{\abehtype(\idvar)}\isdef 1$,
$\height{\tuple{archtype}(\aterm_1, \ldots, \aterm_n)} \isdef
\max\set{\height{\aterm_i} \mid i \in \interv{1}{n}} + 1$ and
$\height{\nu x ~.~ \aterm_1} \isdef \height{\aterm_1}$.  \fi Moreover, for
each behavioral term $\abehterm$, the endpoint of any such chain
starting with $\abehterm$ is unique (modulo commutativity and
associativity of the $\cdot$ and $+$ architecture constructors) and is
denoted by $\canonic{\abehterm}$.

We are now in position to define the semantics of a behavioral term
$\abehterm$, as a (possibly infinite) set of behaviors. Let $\eta$ be
a ground substitution, such that $\fv{\abehterm} \subseteq
\dom(\eta)$, and $\asys$ be a rewriting system. First, we define the
semantics of a well-instantiated ground term in canonical form
$\canonic{\aterm} = \tuple{\archtype}(\aterm_1, \ldots, \aterm_n)$,
from the ground set of $\abehterm$, namely $\aterm \in
\instant{\abehterm}{\eta,\asys}$. Because the flattenning relation is
applied exhaustively to $\aterm$, it must be the case that $\aterm_k =
\abehtype_k(i_k)$, where $i_k \in \ids$, for all $k \in
\interv{1}{n}$.  Then $\sem{\canonic{\aterm}}{}$ is the behavior
$\sem{\archtype}{}(\abehtype_1(i_1), \ldots, \abehtype_n(i_n))$,
defined by (\ref{infrule:arch}). The semantics is lifted from ground
terms to arbitrary behavioral terms:
\begin{equation}\label{eq:sem}
\sem{\abehterm}{\eta,\asys} \isdef \bigcup_{\aterm \in
  \instant{\abehterm}{\eta,\asys}}\sem{\canonic{\aterm}}{}
\end{equation}
We omit writing $\eta$ when $\abehterm$ is closed. \ifLongVersion For
instance, the semantics of the term $\nu x \nu y \nu z ~.~
\mathit{acq}(x) \cdot \mathit{lock}(z) + \mathit{acq}(y) \cdot
\mathit{lock}(z) + \mathit{rel}(x) \cdot \mathit{unlock}(z) +
\mathit{rel}(y) \cdot \mathit{unlock}(z)(\mathit{Task}(x),
\mathit{Task}(y)$, $\mathit{Lock}(z))$ is the set of behaviors that
differ from the one in Fig. \ref{fig:behavioral-types}b only by a
permutation of identifiers. \fi For example, Fig. \ref{fig:ring}d
shows the behavior obtained by the following sequence alternating
rewriting and flattening steps:
\[\begin{array}{l}
\mathit{Ring}() \stackrel{\text{(\ref{rule:ring})}}{\leftarrow} \nu y_1 \nu y_2 ~.~ \tuple{\mathit{out}(y_2) \cdot
\mathit{in}(y_1)} (\mathit{Chain}(y_1,y_2)) 
\stackrel{\text{(\ref{rule:chain1})}}{\leftarrow} \\[-1mm]
\nu y_1 \nu y_2 \nu y^1_1 ~.~ \tuple{\mathit{out}(y_2) \cdot \mathit{in}(y_1) + 
\mathit{out}(y_1) \cdot \mathit{in}(y^1_1)}(\mathit{CType}(y_1), \mathit{Chain}(y^1_1,y_2)) 
\stackrel{\text{(\ref{rule:chain2})}}{\leftarrow} \\[1mm]
\nu y_1 \nu y_2 \nu y^1_1 ~.~ \tuple{\mathit{out}(y_2) \cdot \mathit{in}(y_1) + 
\mathit{out}(y_1) \cdot \mathit{in}(y^1_1) + 
\mathit{out}(y^1_1) \cdot \mathit{in}(y_2)}(\mathit{CType}(y_1), \mathit{CType}(y^1_1), \mathit{CType}(y_2)). 
\end{array}\]

\ifLongVersion
\subsection{Normalized Terms and Rewriting Systems}
\label{sec:normalized}
To ease the upcoming developments, we shall consider closed behavioral
terms and rewriting systems that meet the following:
\begin{assumption}\label{ass:one-inst}
  Each bound variable in $\abehterm$ is instantiated exactly once in
  each predicate-less term $\aterm \in \charterm{\abehterm}{\asys}$.
\end{assumption}
In the rest of this section we shall prove that this assumption loses
no generality. To this end we effectively transform the given system
$\asys$ into an equivalent \emph{normalized} rewriting system
$\asys_N$, yielding exactly those predicate-less terms produced by
$\asys$, in which every variable is instantiated exactly once. The
detailed construction of normalized rewriting systems is given in the
proof of the following:

\begin{proposition}\label{prop:normalized}
For each rewriting system $\asys$ one can effectively construct a
rewriting system $\asys_N$ and a mapping $\Upsilon : \preds
\rightarrow 2^\nat$, such that the following hold: \begin{compactenum}
\item\label{it1:prop:normalized} for each $\apred \in \preds$ and each $i
  \in \interv{1}{\#(\apred)}$, we have $i \in \Upsilon(\apred)$ iff
  $x_i$ is instantiated exactly once in each predicate-less term
  $\aterm \in \charterm{\apred(x_1, \ldots,
    x_{\#(\apred)})}{\asys_N}$.
\item\label{it2:prop:normalized} for each closed behavioral term
  $\abehterm$, we have $\sem{\abehterm}{\asys} =
  \sem{\abehterm}{\asys_N}$.
\end{compactenum} 
Moreover, $\asys_N$ is built in time $\bigO(\size{\asys} \cdot
2^{\width{\asys} \cdot \alpha(\asys)})$, where $\alpha(\asys) \isdef
\max\set{\#(\apred) \mid \apred \in \predset \text{ occurs in }
  \asys}$.
\end{proposition}
\proof{
  The idea is to consider, for each subset
of the set of arguments $I \subseteq \interv{1}{\#(\apred)}$ of a
predicate symbol $\apred \in \predset$, a fresh predicate symbol
$\apred_I$ of arity $\#(\apred_I) = \#(\apred)$, such that
$\set{y_i \mid i \in I}$ is the set of symbols instantiated
exactly once in every rewriting of
$\apred(y_1,\ldots,y_{\#(\apred)})$ by the rules in
$\asys_N$. The rules of $\asys_N$ associated with $\apred_I$ are
defined by the conditions below: \begin{compactenum}
\item\label{it1:normalized} for each rule $\apred(x_1, \ldots,
  x_{\#(\apred)}) \leftarrow_\asys \abehterm$, where
  $\apred^1(y^1_1, \ldots, y^1_{\#(\apred_1)}), \ldots,
  \apred^k(y^k_1, \ldots, y^k_{\#(\apred_k)})$ are the
  predicate subterms of $\abehterm$, there exists zero or more rules
  $\apred_I(x_1, \ldots, x_{\#(\apred)}) \leftarrow_{\asys_N}
  \abehterm'$, where $\abehterm'$ is obtained by replacing each term
  $\apred^\ell(y^\ell_1, \ldots, y^\ell_{\#(\apred_1)})$,
  $\ell \in \interv{1}{k}$ with a predicate term
  $\apred^\ell_{I_\ell}(y^\ell_1, \ldots,
  y^\ell_{\#(\apred_1)})$, such that for all $i \in I$, either
  $x_i \in \ins{\abehterm}$ or there exists $\ell \in \interv{1}{k}$
  such that $x_i = y^\ell_h$, for some $h \in I_\ell$. 
\item\label{it2:normalized} $\apred(x_1, \ldots, x_{\#(\apred)})
  \leftarrow_{\asys_N} \apred_I(x_1, \ldots, x_{\#(\apred)})$, for
  each $I \subseteq \interv{1}{\#(\apred)}$.
\item\label{it3:normalized} for each rule $\apred_I(x_1, \ldots,
  x_{\#(\apred)}) \leftarrow_{\asys_N} \abehterm$, where
  $\apred^1_{I_1}(y^1_1, \ldots, y^1_{\#(\apred_1)}), \ldots,
  \apred^k_{I_k}(y^k_1, \ldots, y^k_{\#(\apred_k)})$ are the predicate
  subterms of $\abehterm$, the following hold, for each symbol $y \in
  \vars \cup \ids$, either one of the following
  applies: \begin{compactenum}[(a)]
    \item\label{it3.1:normalized} $y$ occurs instantiated exactly once
      once in $\abehterm$, or 
    \item\label{it3.2:normalized} there exists exactly one $\ell \in
      \interv{1}{k}$ and one $h \in I_\ell$, such that $y =
      \idvar^\ell_h$.
  \end{compactenum}
\end{compactenum}
Note that the size of and the time required to build $\asys_N$ are
bounded by $\size{\asys} \cdot 2^{\width{\asys} \cdot \alpha(\asys)}$,
where $\alpha(\asys) \isdef \max\set{\#(\apred) \mid \apred \in
  \predset \text{ occurs in } \asys}$. Below we prove the two points
of the statement: 

\noindent(\ref{it1:prop:normalized}) We prove the equivalent statement: 
\[\charterm{\apred(\idvar_1, \ldots,
  \idvar_{\#(\apred)})}{\asys_N} = \set{\aterm \in
  \charterm{\apred(\idvar_1, \ldots, \idvar_{\#(\apred)})}{\asys}
  \mid \idvar_i \text{ is instantiated once in } \aterm,~ i \in
  \interv{1}{\#(\apred)}}\]
``$\subseteq$'' Let $\aterm \in \charterm{\apred(\idvar_1,
    \ldots, \idvar_{\#(\apred)})}{\asys_N}$ be a predicate-less
  term. Then there exists a rewriting tree $T \in
  \rtrees{\apred(\idvar_1, \ldots, \idvar_{\#(\apred)})}{\asys_N}$
  such that $\aterm = \charterm{T}{}$. Since the only rules defining
  $\apred(x_1, \ldots, x_{\#(\apred)})$ in $\asys_N$ are of the form
  $\apred(x_1, \ldots, x_{\#(\apred)}) \leftarrow_{\asys_N}
  \apred_I(x_1, \ldots, x_{\#(\apred)})$, by point
  (\ref{it2:normalized}) of the definition of $\asys_N$, it must be
  the case that the root of $T$ has only one child and let
  $\proj{T}{1}$ be the subtree of $T$ rooted in the single node below
  the root, for some $I \subseteq \interv{1}{\#(\apred)}$. Then we
  build a rewriting tree $U \in \rtrees{\apred(\idvar_1, \ldots,
    \idvar_{\#(\apred)})}{\asys}$ isomorphic with $\proj{T}{1}$, by
  deleting the $J$ subscript from each predicate symbol $\apred'_J$
  that occurs in $\proj{T}{1}$. It is easy to see that the result is
  indeed a rewriting tree in $\asys$, by the point
  (\ref{it1:normalized}) of the definition of $\asys_N$. Suppose, for
  a contradiction, that $\idvar_i$ occurs instantiated at least twice
  in $\aterm$, for some $i \in \interv{1}{\#(\apred)}$ (case where
  $\idvar_i$ is not instantiated is left to the reader). Two cases are
  possible: \begin{compactenum}[a.]
  \item There exists $w \in \nodes(U)$ such that $U(w) = \left(
    \apred^w_{I_w}(\idvar^w_1,\ldots,\idvar^w_{\#(\apred^w_{I_w})})
    \leftarrow \abehterm^w\right)$, such that $\idvar_i$ occurs
    instantiated twice in $\abehterm^w$, which contradicts point
    (\ref{it3.1:normalized}) from the definition of $\asys_N$.
  \item There exist $w_1 \neq w_2 \in \nodes(U)$, such that $U(w_j) =
    \left(\apred^{w_j}_{I_{w_j}}(\idvar^{w_j}_1, \ldots,
    \idvar^{w_j}_{\#(\apred^{w_j}_{I_{w_j}})}) \leftarrow
    \abehterm^{w_j}\right)$, $j = 1,2$ and $\idvar_i \in
    \ins{\abehterm^{w_1}} \cap \ins{\abehterm^{w_2}}$. Let $w$ be the
    largest common prefix of $w_1$ and $w_2$ and let $U(w) =
    \left(\apred^w_{I_w}(\idvar^w_1,\ldots,\idvar^w_{\#(\apred^w_{I_w})})
    \leftarrow \abehterm^w\right)$. Then one of the following
    applies: \begin{compactitem}
    \item there exists a predicate subterm $\apred^1_{I_1}(\idvar^1_1,
      \ldots, \idvar^1_{\#(\apred^1)})$ of $\abehterm^w$, such that
      $\idvar_i = \idvar^1_{\ell_1} = \idvar^1_{\ell_2}$, for some
      $\ell_1 \neq \ell_2 \in I_1$, which contradicts point
      (\ref{it3.2:normalized}) from the definition of $\asys_N$.
    \item there exist two predicate subterms
      $\apred^j_{I_j}(\idvar^j_1, \ldots,
      \idvar^j_{\#(\apred^j_{I_j})})$ of $\abehterm^w$, $j=1,2$, such
      that $\idvar_i = \idvar^1_{\ell_1} = \idvar^2_{\ell_2}$, for
      some $\ell_j \in I_j$, $j = 1,2$, which is again in
      contradiction with point (\ref{it3.2:normalized}) from the
      definition of $\asys_N$.
    \end{compactitem}
  \end{compactenum}

  \noindent``$\supseteq$'' Let $\aterm \in \charterm{\apred(\idvar_1,
    \ldots, \idvar_{\#(\apred)})}{\asys}$ be a predicate-less term,
  such that $\idvar_i$ is instantiated once in $\aterm$, for each $i
  \in \interv{1}{\#(\apred)}$. Then there exists a rewriting tree $T
  \in \rtrees{\apred(\idvar_1, \ldots, \idvar_{\#(\apred)})}{\asys}$,
  such that $\aterm = \charterm{T}{}$. We build a rewriting tree $U$,
  such that $\nodes(U) = \set{1w \mid w \in \nodes(T)} \cup
  \set{\lambda}$ and the labels of $U$ are defined bottom-up as
  follows: \begin{compactitem}
  \item if $w \in \leaves(T)$ and $T(w) =
    \left(\apred^w(\idvar^w_1,\ldots,\idvar^w_{\#(\apred^w)})
    \leftarrow \abehterm^w\right)$, then we define $U(1w) \isdef
    \left(\apred^w_{I_w}(\idvar^w_1,\ldots,\idvar^w_{\#(\apred^w)})
    \leftarrow \abehterm^w\right)$, where $I_w \isdef \{i \in
    \interv{1}{\#(\apred^w)} \mid \idvar^w_i \in \ins{\abehterm^w}\}$.
  \item else, if $w \in \nodes(T) \setminus \leaves(T)$ let $T(w) =
    \left(\apred^w(\idvar^w_1,\ldots,\idvar^w_{\#(\apred^w)})
    \leftarrow \abehterm^w\right)$ and $w1, \ldots, wk$ be the
    children of $w$ in $T$. Moreover, $U(1w\ell) =
    \left(\apred^{w\ell}_{I_{w\ell}}(\idvar^{w\ell}_1,\ldots,\idvar^{w\ell}_{\#(\apred^{w\ell})})
    \leftarrow \abehterm^{w\ell}\right)$ has already been defined, for
    all $\ell \in \interv{1}{k}$. Then we define $U(1w) \isdef
    \left(\apred^w_{I_w}(\idvar^w_1,\ldots,\idvar^w_{\#(\apred^w)})
    \leftarrow \abehterm^w\right)$, where $I_w \isdef \{i \in
    \interv{1}{\#(\apred)} \mid \idvar^w_i \in \ins{\abehterm^w}
    \text{ or } \exists \ell \in \interv{1}{k} ~.~ \idvar^w_i =
    \idvar^{w\ell}_h \text{ and } h \in I_{w\ell}\}$.
  \item $U(\lambda) \isdef \left(\apred(\idvar_1, \ldots,
    \idvar_{\#(\apred)}) \leftarrow
    \apred_I(\idvar_1,\ldots,\idvar_{\#(\apred)})\right)$, where $U(1)
    = \left(\apred_I(\idvar_1,\ldots,\idvar_{\#(\apred^1)}) \leftarrow
    \abehterm\right)$ has already been defined and $\#(\apred) =
    \#(\apred^1)$.
  \end{compactitem}
  It is easy to check that indeed $U \in \rtrees{\apred(\idvar_1,
    \ldots, \idvar_{\#(\apred)}), \abehterm}{\asys_N}$ (Definition
  \ref{def:rewriting-tree}).

\noindent(\ref{it2:prop:normalized}) We prove the statement in general,
when $\abehterm$ is not necessarily closed and $\eta$ is a ground
substitution such that $\fv{\abehterm} \subseteq \dom(\eta)$. Because
the set $I$ of parameters instantiated in every $\asys_N$-rewriting of
a predicate term $\apred_I(\idvar_1, \ldots, \idvar_{\#(\apred)})$ is
known \`a~priori, we consider w.l.o.g, for an arbitrary behavioral
term $\abehterm$, that $\charterm{\abehterm}{\asys_N}$ is the set of
predicate-less terms obtained by a rewriting of $\abehterm$, in which
every variable/identifier occurs instantiated exactly once.  We
compute:
\[\begin{array}{rclr}
\sem{\abehterm}{\eta,\asys_N} & = & 
\bigcup_{\aterm \in \instant{\abehterm}{\eta,\asys_N}} \sem{\canonic{\aterm}}{} \\ 
& = & \bigcup_{\uterm \in \charterm{\abehterm}{\asys_N}} \bigcup_{\aterm \in \instant{\uterm}{\eta}} \sem{\canonic{\aterm}}{} \\
& = & \bigcup_{\begin{array}{l}
    \scriptstyle{\uterm \in \charterm{\abehterm}{\asys}} \\[-2mm] 
    \scriptstyle{\text{each symbol is instantiated exactly once in $\uterm$}}
\end{array}} 
\bigcup_{\aterm \in \instant{\uterm}{\eta}} \sem{\canonic{\aterm}}{} 
& (\text{by point (\ref{it1:prop:normalized})}) \\
& = & \bigcup_{\uterm \in \charterm{\abehterm}{\asys}} \bigcup_{\aterm \in \instant{\uterm}{\eta}} \sem{\canonic{\aterm}}{}
& (\text{definition of $\instant{\uterm}{\eta}$}) \\
& = & \sem{\abehterm}{\eta,\asys} & \text{\qed}
\end{array}\]}

Given a closed behavioral term $\abehterm$, we assume first that every
(necessarily bound) variable $x$ of $\abehterm$ occurs either in
exactly one instance atom $\abehtype(x)$ or in exactly one predicate
atom $\apred(x_1, \ldots, x_{\#\apred})$ of $\abehterm$ as $x = x_i$,
for some $i \in \Upsilon(\apred)$, where $\Upsilon$ is the mapping
associated with $\asys_N$ in Proposition \ref{prop:normalized}. This
assumption is w.l.o.g. because, if $x$ occurs in two or more (instance
of predicate) atoms in violation of the above condition, $\abehterm$
has no associated behaviors, i.e.\ $\sem{\abehterm}{\asys_N} =
\emptyset$, implying that $\sem{\abehterm}{\asys} = \emptyset$, by
Proposition \ref{prop:normalized}. Moreover, if $x$ is never
instantiated in $\abehterm$, the interactions involving some port $p(x)$ 

Second, we assume that, in each subterm of $\abehterm$ of the form
$\tuple{\archtype}(\abehterm_1, \ldots, \abehterm_n)$ at most one of the terms
$\abehterm_1, \ldots, \abehterm_n$ is an instantiation atom, the rest
being predicate atoms. Again, this assumption loses no generality,
because every subterm $\abehtype(\idvar)$ can be replaced with a fresh
predicate atom $\apred_{\abehtype}(\idvar)$, by adding the rule
$\apred_{\abehtype}(x) \leftarrow \abehtype(x)$ to $\asys$ (see rule
(\ref{rule:tll5}) from Example \ref{ex:tll}). The purpose of this
assumption is to be able to identify indices of instances with the
nodes of a rewriting tree (a detailed explanation will be given in
\S\ref{sec:behterm-parametric}). The behavioral terms satisfying the
above conditions are said to be \emph{normalized} as well.\fi

\section{The Parametric Safety Problem}
\label{sec:safety}

Having defined a language for specification of architectures, we move
on to the problem of verifying that every behavior generated by a
rewriting system, starting with a given behavioral term, is safe with
respect to a set of error configurations. This problem is challenging,
because we ask for a proof of safety that holds \emph{for every}
ground instantiation of some predicate-less rewriting of the
behavioral term.

Intuitively, a set of behaviors is said to be \emph{parametric} if
each behavior in the set is obtained from the same pattern, by
assigning different values to several designated variables, called
\emph{parameters}. Formally, a \emph{parametric system} is a tuple
$\mathcal{C} = \tuple{\abehtype_1, \ldots, \abehtype_K, \aarch}$,
where $\abehtype_i \in \behtypeset$ are component types and $\aarch$
maps a tuple $\vec{T} = \tuple{T_1, \ldots, T_N}$ of sets of
identifiers $T_1, \ldots, T_N \subseteq\ids$, to an architecture,
denoted as $\aarch(\vec{T})$. Intuitively, the tuple of sets $\vec{T}$
is a \emph{structural parameter} of the system, that
defines\begin{inparaenum}[(i)] 
\item the architecture which coordinates the instances of
  $\abehtype_1, \ldots, \abehtype_K$ and
\item the set of instances belonging to each behavior type. 
\end{inparaenum}
For presentation purposes, we defer the precise definitions to
\S\ref{sec:behterm-parametric}.
The behavior resulting from the application, using the composition
rule (\ref{infrule:arch}), of the architecture $\aarch(\vec{T})$ to
these instances is denoted as $\mathcal{C}(\vec{T})$.




The \emph{parametric safety problem} asks whether each behavior
$\mathcal{C}(\vec{T})$ of a parametric system $\mathcal{C}$ is safe
w.r.t. a given set $\errstates$ of configurations. Since, in general,
the parametric safety problem is undecidable, we resort to a sound but
necessarily incomplete solution, that consists in computing
\emph{safety invariants}. Given a behavior $\abeh$, an invariant
$\ainv$ of $\abeh$ is a superset of the set of reachable
configurations of $\abeh$, thus $\abeh$ is safe w.r.t. $\errstates$ if
$\ainv \cap \errstates = \emptyset$ (the reversed implication is
clearly not true in general). Since we consider a parametric system,
the challenge is computing a parametric safety invariant, i.e.\ a
pattern that defines an invariant for each behavior
$\mathcal{C}(\vec{T})$, determined by a choice of $\vec{T}$.

In contrast with the classical approach to invariant synthesis based
on a fixpoint iteration in an abstract domain \cite{CousotCousot79},
we focus on a particular class of invariants that can be obtained
directly from the description of the parametric system. These
invariants are called \emph{structural} in the following. The
structural invariants considered in this paper are mostly inspired by
the following notions:
\begin{definition}\label{def:trap}
A \emph{trap} $\theta$ of a behavior $\abeh = \tuple{\iports,
  \istates, \iinitstate, \irules}$ is a subset of $\istates$ such
that, for any two configurations $\sigma$ and $\sigma'$ of $\abeh$,
such that $\sigma \arrow{}{\abeh} \sigma'$, we have $\sigma \cap
\theta \neq \emptyset$ only if $\sigma' \cap \theta \neq \emptyset$. A
trap $\theta$ is \emph{marked} iff $\theta \cap \iinitstate \neq
\emptyset$. The \emph{trap invariant} of $\abeh$ is the set
$\Theta(\abeh) \isdef \set{\sigma \subseteq \istates \mid \sigma \cap
  \theta \neq \emptyset, \text{ for each marked trap } \theta \text{
    of } \abeh}$.
\end{definition}
To understand why $\Theta(\abeh)$ is an invariant of $\abeh$, note
that $\Theta(\abeh)$ contains the initial configuration of $\abeh$ and
is closed under the transition relation $\arrow{}{\abeh}$. Since the
set of reachable configurations of $\abeh$ is the smallest such set,
it follows that $\Theta(\abeh)$ is an over-approximation of the
reachable configurations of $\abeh$, hence an invariant. 

\subsection{The Weak Sequential Calculus of $\kappa$ Successors}
\label{sec:wssk}

The structural invariants and the sets of unsafe configurations will
be described using a restriction of monadic second order logic (\mso)
to trees of branching $\kappa$, where $\kappa>0$ is an integer
constant. Let $\Vars = \set{X,Y,Z,\ldots}$ be a countably infinite set
of second order variables. The formul{\ae} of \wss{\kappa} are defined
by the following syntax:
\[\begin{array}{rclr}
\tau & ::= & \wssroot \mid x \in \vars \mid \succ_i(\tau_1), ~i\in\interv{0}{\kappa-1} & \text{ terms} \\
\phi & ::= & \tau_1 = \tau_2 \mid X(\tau) \mid \phi_1 \wedge \phi_2 \mid \neg\phi_1 
\mid \exists x ~.~ \phi_1 \mid \exists X ~.~ \phi_1 & \text{ formul{\ae}} \\
\end{array}\]
As usual, we write $\phi_1 \vee \phi_2 \isdef \neg(\neg\phi_1 \wedge
\neg\phi_2)$, $\phi_1 \rightarrow \phi_2 \isdef \neg\phi_1 \vee
\phi_2$, $\phi_1 \leftrightarrow \phi_2 \isdef \phi_1 \rightarrow
\phi_2 \wedge \phi_2 \rightarrow \phi_1$, $\forall x ~.~ \phi \isdef
\neg\exists x ~.~ \neg\phi$ and $\forall X ~.~ \phi \isdef \neg\exists
X ~.~ \neg\phi$.

\wss{\kappa} formul{\ae} are interpreted over an infinite $\kappa$-ary
tree with nodes $\interv{0}{\kappa-1}^*$, where first order variables
$x \in \vars$ range over individual nodes $n \in
\interv{0}{\kappa-1}^*$, second order variables $X \in \Vars$ range
over finite sets of nodes $T \subseteq \interv{0}{\kappa-1}^*$,
$\wssroot$ is a constant symbol interpreted as $\epsilon$ and, for all
$i \in \interv{0}{\kappa-1}$, the function symbol $\succ_i$ is
interpreted by the total function $n \mapsto ni$. Given a valuation
$\nu : \vars \cup \Vars \rightarrow \interv{0}{\kappa-1}^* \cup
2^{\interv{0}{\kappa-1}^*}$, such that $\nu(x) \in
\interv{0}{\kappa-1}^*$, for each $x \in \vars$ and $\nu(X) \subseteq
\interv{0}{\kappa-1}^*$, for each $X \in \Vars$, the satisfaction
relation $\nu \models \phi$ is defined inductively on the structure of
the formula $\phi$\ifLongVersion:
\[\begin{array}{rclcl}
\nu & \models & \tau_1 = \tau_2 & \iff & \nu({\tau_1}) = \nu({\tau_2}) \\
\nu & \models & X(\tau) & \iff & \nu(\tau) \in \nu(X) \\
\nu & \models & \exists x ~.~ \phi_1 & \iff &
\nu[x\leftarrow w] \models \phi_1 \text{, for some node } w \in \interv{0}{\kappa-1}^* \\
\nu & \models & \exists X ~.~ \phi_1 & \iff &
\nu[X\leftarrow W] \models \phi_1 \text{, for some finite set } W \subseteq \interv{0}{\kappa-1}^*
\end{array}\]
where $\nu(\tau)$ is the homomorphic extension of $\nu$ to the term
$\tau$ and $\nu[x \leftarrow w]$ ($\nu[X \leftarrow W]$) is the
valuation that acts like $\nu$, except for $x$ ($X$) which is mapped
to $w$ ($W$). The meaning of the boolean connectives is the usual
one.\fi A valuation $\nu$ is a \emph{model} of a
formula $\phi$ if and only if $\nu \models \phi$. A formula is
\emph{satisfiable} if and only if it has a model.

\subsection{Parametric Systems Defined by Behavioral Terms}
\label{sec:behterm-parametric}

We define the parametric component-based system $\mathcal{C} =
\tuple{\abehtype_1, \ldots, \abehtype_K, \aarch}$ corresponding to a
given closed behavioral term $\abehterm$ and a rewriting system
$\asys$. \ifLongVersion Without loss of generality, we consider that
$\abehterm$ and $\asys$ are normalized (\S\ref{sec:normalized}).
\fi This allows us to identify
the indices of instances with the nodes of a rewriting tree
(Definition \ref{def:rewriting-tree}), in order to describe parametric
invariants using \wss{\kappa}. More precisely, we identify the index
of a component instantiated by an atom $\abehtype(x)$ of $\abehterm$,
with the unique node of the rewriting tree $\mathcal{T} \in
\rtrees{\abehterm}{\asys}$ labeled by that atom. Note that, by
Assumption \ref{ass:one-inst}, the index of the $\abehtype(x)$
component is uniquely determined by $\mathcal{T}$. Consequently, in
the rest of the paper, we shall silently identify $\ids$ with
$\interv{0}{\kappa-1}^*$.

In principle, by fixing a particular interpretation of indices in a
ground term $\aterm \in \instant{\abehterm}{\asys}$, we also restrict
the set of behaviors considered, i.e.\ we consider a strict subset of
$\sem{\abehterm}{\asys}$ (\ref{eq:sem}). This particular restriction
is, however, without consequences for the soundness of the
verification method, because ground terms that differ only by a
permutation of indices generate behaviors that are bisimilar and have
the same safety properties (modulo a permutation of indices). We shall
silently assume, from now on, that the set of unsafe configurations
$\errstates$ from the specification of a parametric safety problem is
closed under permutations of indices. This is the case when the
\wss{\kappa} definition of $\errstates$ does not use successor
functions and only compares first order variables for equality
(e.g.\ \ref{eq:deadlock}).

\begin{figure}[t!]
  \caption{Encoding Rewriting Trees, Instance Sets and Configurations in \wss{\kappa}}  
  \label{fig:encoding}
  \vspace*{-.5\baselineskip}
\[\begin{array}{rcl}
\mathit{RTree}(\vec{U}) & \isdef & 
\forall x ~.~ \bigwedge_{1 \leq i < j \leq N} \Big(\neg U_i(x) \vee \neg U_j(x)\Big) \wedge 
U_1(x) \leftrightarrow x = \wssroot ~\wedge \\
&& \forall x ~.~ \bigwedge_{\arule_i \in \asys} \bigwedge_{\ell=0}^{\kappa-1} 
U_i(\succ_\ell(x)) \rightarrow \bigvee_{\arule_j \in \asys} U_j(x) ~\wedge \\
&& \forall x ~.~ \bigwedge_{\scriptstyle{\arule_i = \left(\apred'(x_1, \ldots, x_{\#(\apred')}) \leftarrow_\asys \abehterm'\right)}} 
\bigwedge_{j=0}^{\npred{\abehterm'}-1}
U_i(x) \rightarrow \\ 
&& \hspace*{4cm}\Big(\bigvee_{\begin{array}{l}
    \scriptstyle{\arule_\ell = \left(\apred''(x_1, \ldots, x_{\#(\apred'')}) \leftarrow_\asys \abehterm''\right)} \\[-.5mm]
    \scriptstyle{\apred''(\idvar_1, \ldots, \idvar_{\#(\apred'')}) = \pred{j}{\abehterm'}}
\end{array}} U_{\ell}(\succ_j(x))\Big)
\\[2mm]
\mathit{Inst}(\vec{U}, \vec{Z}) & \isdef & 
\forall x .\bigwedge_{i=1}^K Z_i(x) \leftrightarrow 
\bigvee_{\begin{array}{l}
    \scriptstyle{\arule_j = \left(\apred'(x_1, \ldots, x_{\#(\apred')}) \leftarrow_{\asys_\abehterm} \abehterm'\right)} \\[-.5mm]
    \scriptstyle{\abehtype_i(z) \in \instances{\abehterm'}}
\end{array}} U_j(x) 
\\[2mm]
\mathit{Config}(\vec{X}, \vec{Z}) & \isdef & \forall x .
\bigwedge_{S \neq T \in \bigcup_{j=1}^K \states_{\abehtype_j}} \Big(\neg X_S(x) \vee \neg X_T(x)\Big) \wedge 
\Big(\bigvee_{S \in \bigcup_{j=1}^K \states_{\abehtype_j}} X_S(x)\Big) \leftrightarrow \Big(\bigvee_{j=1}^K Z_j(x)\Big) 
\end{array}\]
\vspace*{-2\baselineskip}
\end{figure}

Let us consider that $\asys_\abehterm = \asys \cup (\apred_\abehterm()
\leftarrow \abehterm)$ consists of the rules $\arule_1, \ldots,
\arule_N$, such that $\arule_1 = (\apred_\abehterm() \leftarrow
\abehterm)$. We use a designated tuple of second order variables
$\vec{U} = \tuple{U_1, \ldots, U_N}$, where each variable $U_i$ is
interpreted as the set of tree nodes labeled with the rule $\arule_i$
in the rewriting tree. Note that, with this convention, $U_{1}$ is a
singleton containing the root of the rewriting tree (Definition
\ref{def:rewriting-tree}). We say that a tuple of sets of identifiers
$\vec{T} = \tuple{T_1, \ldots, T_N}$ is \emph{parameter-compatible
  with $\asys$ and $\abehterm$} iff any valuation $\nu$, such that
$\nu(U_i) = T_i$, for all $i \in \interv{1}{N}$, is a model of the
$\mathit{RTree}(\vec{U})$ formula (Fig. \ref{fig:encoding}).
Note that this formula is a \wss{\kappa} encoding of the conditions
from Definition \ref{def:rewriting-tree}. The above formul{\ae} depend
implicitly on $\asys$ and $\abehterm$, which will be silently assumed
in the following.

We are now in position to define the parametric system $\mathcal{C} =
\tuple{\abehtype_1, \ldots, \abehtype_K, \aarch}$, corresponding to
$\asys$ and $\abehterm$. First, let $\abehtype_1, \ldots, \abehtype_K$
be the component types that occur in $\abehterm$ and in the rules of
$\asys$. Second, we define $\mathcal{A}$ as a partial mapping of the
sets $T_1, \ldots, T_N \subseteq \ids$ to an architecture defined
whenever $\vec{T} = \tuple{T_1, \ldots, T_N}$ is parameter-compatible
with $\asys$ and $\abehterm$. Since, in this case, we have $[U_1
  \leftarrow T_1, \ldots, U_N \leftarrow T_N] \models
\mathit{RTree}(\vec{U})$, the sets $T_1, \ldots, T_N$ uniquely
determine a rewriting tree $\mathcal{T} \in
\rtrees{\abehterm}{\asys}$, such that $T_i \subseteq
\nodes(\mathcal{T})$ is the set of nodes labeled by the rule
$\arule_i$, for all $i \in \interv{1}{N}$.

Further, let $\aterm \in \instant{\charterm{\mathcal{T}}{}}{}$ be
unique ground term defined in the following way: for each instance
atom $\abehtype_i(x)$ that occurs in $\abehterm$, the variable $x$ is
substituted with the unique node of $\mathcal{T}$ where this atom
occurs. This substitution determines the sets of instances for each
behavioral type $\abehtype_1, \ldots, \abehtype_K$, encoded by the
second order variables $\vec{Z} = \tuple{Z_1, \ldots, Z_K}$, in the
$\mathit{Inst}(\vec{U}, \vec{Z})$ formula (Fig. \ref{fig:encoding}).
Note that, by Assumption \ref{ass:one-inst}, there is at most one node
$w \in \nodes(\mathcal{T})$ such that $\mathcal{T}(w) =
\left(\apred_w(x_1, \ldots, x_{\#(\apred_w)}) \leftarrow
\abehterm_W\right)$ and $\abehtype_i(x) \in
\instances{\abehterm}$. Moreover, each such node contains at most one
instance atom, thus different instance atoms are assigned different
identifiers. Finally, the architecture $\aarch(\vec{T})$ is the union
of the ground architectures that occur in $\aterm$, formally
$\aarch(\vec{T}) \isdef \sem{\archtype}{}$, where $\canonic{\aterm} =
\tuple{\archtype}(\aterm_1, \ldots, \aterm_n)$ is the canonical form
of $\aterm$ obtained by exhaustive application of the flattening
relation (\ref{eq:arch-sum}).

\subsection{Trap Invariants for Behavioral Terms}
\label{sec:trap-inv}

Let $\mathcal{C} = \tuple{\abehtype_1, \ldots, \abehtype_K, \aarch}$
be the parametric system corresponding to the given behavioral term
$\abehterm$ and the rewriting system $\asys$. The sets of
configurations of $\mathcal{C}$ are represented by tuples of second
order variables $\vec{X} \isdef \tuple{X_S \mid S \in \bigcup_{j=1}^K
  \states_{\abehtype_j}}$ and $\vec{Y} \isdef \tuple{Y_S \mid S \in
  \bigcup_{j=1}^K \states_{\abehtype_j}}$, where a variable $X_S$
(respectively $Y_S$) encodes the set of indices $i \in \ids$ such that
the instance $\abehtype_j(i)$ is in state $S(i)$, for all $j \in
\interv{1}{K}$. For a mapping $\nu : \vec{X} \rightarrow 2^\ids$, we
define $\nu(\vec{X}) \isdef \tuple{\nu(X_S) \mid S \in \bigcup_{j=1}^K
  \states_{\abehtype_j}}$. The $\mathit{Config}$ formula
(Fig. \ref{fig:encoding}) ensures that $\nu(\vec{X})$ defines a
configuration $\sigma$, for each satisfying valuation $\nu$, by
requiring that the sets assigned to $\vec{X}$ are a partition of the
set of indices of the instances from the system, assigned to
$\vec{Z}$. If $\nu \models \mathit{Config}(\vec{X}, \vec{Z})$, we
write \(\nu(\vec{X}) \rhd \sigma$ iff $\sigma = \{S(i) \mid S \in
\states_{\abehtype_j},~ i \in \nu(X_S),~ j \in \interv{1}{K}\}\).


For the time being, we assume the existence of a \wss{\kappa}\ formula
satisfying the condition below, the definition of which will be given
in \S\ref{sec:trap-inv}:
\begin{eqnarray}\label{eq:flow-cond}
\nu \models \mathit{Flow}(\vec{X},\vec{Y},\vec{U}) & \iff & 
\text{ $\nu(\vec{X}) \rhd \pre{\aint}$ and $\nu(\vec{Y}) \rhd \post{\aint}$, for some }
\aint \in \aarch(\nu(\vec{U})) 
\end{eqnarray}
Intuitively, $\mathit{Flow}$ is satisfied by any valuation that
assigns $\vec{X}$ and $\vec{Y}$ sets of identifiers defining the pre-
and post-configurations of an interaction from the architecture
defined by the valuation of $\vec{U}$. With these definitions, the
following formula translates the conditions of Definition
\ref{def:trap}, describing (parametric) traps:
\[\begin{array}{rcl}
\mathit{Trap}(\vec{X},\vec{U}) & \isdef & \forall \vec{Y}^1 \forall \vec{Y}^2 ~.~ 
\mathit{Flow}(\vec{Y}^1,\vec{Y}^2,\vec{U}) \wedge \mathit{inter}(\vec{X},\vec{Y}^1) 
\rightarrow \mathit{inter}(\vec{X},\vec{Y}^2) \\ 
\mathit{inter}(\vec{X},\vec{Y}) & \isdef & \exists x .
\bigvee_{j=1}^K \bigvee_{S \in \states_{\abehtype_j}} X_S(x) \wedge Y_S(x)
\end{array}\]
where $\vec{Y}^i$ is the copy of the tuple $\vec{Y}$ with variables
superscripted by $i$, for $i=1,2$. The set of configurations defined
by the formula below is the trap invariant (Definition \ref{def:trap})
of $\mathcal{C}$, for each parameter-compatible
interpretation of $\vec{U}$:
\begin{eqnarray}\label{eq:trap-inv}
\mathit{TrapInv}(\vec{X},\vec{U}) & \isdef &
\exists \vec{Z} ~.~ \mathit{Inst}(\vec{U},\vec{Z}) \wedge \mathit{Config}(\vec{X},\vec{Z}) ~\wedge \\
&& \forall \vec{Y}^1 \forall \vec{Y}^2 ~.~ \mathit{Init}(\vec{Y}^1,\vec{Z}) \wedge 
\mathit{Trap}(\vec{Y}^2,\vec{U}) \wedge \mathit{inter}(\vec{Y}^1,\vec{Y}^2) \rightarrow 
\mathit{inter}(\vec{X},\vec{Y}^2) \nonumber \\
\mathit{Init}(\vec{X}, \vec{Z}) & \isdef & \bigwedge_{j=1}^K \forall x ~.~ 
Z_j(x) \leftrightarrow X_{\initstate_{\abehtype_j}}(x) \nonumber
\end{eqnarray}
where the formula $\mathit{Init}$ defines the initial configuration of
the parametric system, in which each instance is in the initial state
of its component type. The following lemma proves that, assuming the
existence of a formula $\mathit{Flow}$ satisfying the condition
(\ref{eq:flow}), the formula $\mathit{TrapInv}$ correctly defines the
(parametric) trap invariant of the parametric system corresponding to
$\asys$ and $\abehterm$:

\begin{lemma}\label{lemma:trap-invariant}
  Let $T_1, \ldots, T_N \subseteq \ids$ be finite sets such that
  $[U_1 \leftarrow T_1, \ldots, U_N \leftarrow T_N] \models
  \mathit{RTree}(\vec{U})$. Then
  $\Theta(\mathcal{C}(\vec{T})) = \{\sigma \mid
  \nu(\vec{X}) \rhd \sigma,~ \nu[U_1 \leftarrow T_1, \ldots, U_N
    \leftarrow T_N] \models \mathit{TrapInv}(\vec{X}, \vec{U})\}$.
\end{lemma}
\proof{ 
  Let $\mathcal{C}_{\asys, \abehterm} = \tuple{\abehtype_1, \ldots,
    \abehtype_K, \aarch}$ be the parametric system corresponding to
  $\asys$ and $\abehterm$. Since $[U_1 \leftarrow T_1, \ldots, U_N
    \leftarrow T_N] \models \mathit{RTree}(\vec{U})$, it is easy to
  prove that there exists a unique rewriting tree $\mathcal{T} \in
  \rtrees{\abehterm}{\asys}$, such that $U_1, \ldots, U_N$ form a
  partition of $\nodes(\mathcal{T})$ and each node in $U_i$ is labeled
  with the rule $\arule_i$ from $\asys_\abehterm = \{\arule_1, \ldots,
  \arule_N\}$. By Assumption \ref{ass:one-inst}, each (necessarily
  bound) variable in $\abehterm$ is instantiated exactly once in
  $\charterm{\mathcal{T}}{}$ and let $P_1, \ldots, P_K \subseteq
  \nodes(\mathcal{T})$ be the sets of nodes such that $P_i$ contains
  those nodes of $\mathcal{T}$ in which an instance of $\abehtype_i$
  is created, for all $i \in \interv{1}{K}$. 

  \noindent''$\subseteq$'' Let $\sigma \in
  \Theta(\mathcal{C})$ be a configuration and $\nu$
  be a valuation such that $\nu(\vec{X}) \rhd \sigma$ and $\nu(Z_i) =
  P_i$, for all $i \in \interv{1}{K}$. Then $\nu \models
  \mathit{Inst}(\vec{U}, \vec{Z}) \wedge \mathit{Config}(\vec{X},
  \vec{Z})$ follows from the choice of $\nu$. Let $\vec{V}^i =
  \tuple{V^i_S \subseteq \nodes(\mathcal{T}) \mid S \in
    \bigcup_{j=1}^K \states_{\abehtype_j}}$, for $i = 1,2$, be tuples
  of sets such that
  \[\mu \models \mathit{Init}(\vec{Y}^1, \vec{Z}) \wedge 
  \mathit{Trap}(\vec{Y}^2, \vec{U}) \wedge \mathit{inter}(\vec{Y}^1,
  \vec{Y}^2)\] where $\mu$ is any extension of $\nu$ that assigns each
  second order variable $Y^i_S$ the set $V^i_S$. Moreover, let
  $\theta^i$ be sets of states, such that $\mu(\vec{Y}^i) \rhd
  \theta^i$, for $i = 1,2$. It is easy to check that: \begin{compactitem}
  \item $\sigma^1$ is an initial configuration of $\mathcal{C}_{\asys,
    \abehterm}$, because $\mu \models \mathit{Init}(\vec{Y}^1,
    \vec{Z})$,
  \item $\sigma^2$ is a trap of $\mathcal{C}_{\asys, \abehterm}$,
    because $\mu \models \mathit{Trap}(\vec{Y}^2, \vec{Z})$, and
  \item $\sigma^1 \cap \sigma^2 \neq \emptyset$, because $\mu \models
    \mathit{inter}(\vec{Y}^1, \vec{Y}^2)$.
  \end{compactitem}
  Then $\sigma^2$ is a marked trap of $\mathcal{C}$.
  By Definition \ref{def:trap}, $\sigma$ intersects with every marked
  trap of $\mathcal{C}_{\asys, \abehterm}$, hence $\sigma \cap
  \sigma^2 \neq \emptyset$, leading to $\mu \models
  \mathit{inter}(\vec{X}, \vec{Y}^2)$, hence $\nu \models
  \mathit{TrapInv}(\vec{X}, \vec{U})$. The ``$\supseteq$'' direction
  follows a similar argument and is left to the reader. \qed}

Assuming that the $\errstates$ set is encoded by a formula
$\mathit{Bad}$, the parametric safety problem has a positive answer if
the following formula is unsatisfiable:
\begin{equation}\label{eq:safety}
\mathit{Safe}(\vec{U}) \isdef \mathit{RTree}(\vec{U}) \wedge \exists \vec{X} ~.~ 
\mathit{TrapInv}(\vec{X},\vec{U}) \wedge \mathit{Bad}(\vec{X},\vec{U})
\end{equation}
As a typical example of a set of unsafe states, we consider the
following definition of deadlock configurations, i.e.\ configurations
in which no interaction can be fired:
\begin{equation}\label{eq:deadlock}
\mathit{DeadLock}(\vec{X},\vec{U}) \isdef \forall \vec{Y}^1 \forall \vec{Y}^2 ~.~ 
\mathit{Flow}(\vec{Y}^1,\vec{Y}^2,\vec{U}) \rightarrow \exists
x . \bigvee_{j=1}^K\bigvee_{S \in \states_{\abehtype_j}} Y^1_S(x)
\wedge \neg X_S(x)
\end{equation}
\ifLongVersion
Note that the set of deadlock configurations
defined by $\mathit{DeadLock}$ is invariant under permutations of
indices.\fi

\subsection{The Flow of a Behavioral Term}
\label{sec:flow}

To complete the definition of trap invariants using \wss{\kappa}, we
are left with defining the $\mathit{Flow}(\vec{X}, \vec{Y}, \vec{U})$
formula (\ref{eq:flow-cond}), that holds whenever $(\vec{X},\vec{Y})$
encodes the pairs of pre- and post-configurations of some interaction
from $\mathcal{C}(\vec{T})$, when $\vec{U}$ are interpreted by the
sets of identifiers $\vec{T}$. We recall that $\asys_{\abehterm} =
\set{\arule_1, \ldots, \arule_N}$ and that we have assumed the rules
in $\asys_\abehterm$ to be of the form \(\apred(x_1, \ldots,
x_{\#(\apred)}) \leftarrow \nu y_1 \ldots \nu y_m ~.\)
\(\tuple{\archtype}(\aterm_1, \ldots, \aterm_n)\), where each
$\aterm_i$ is an atom and at most one $\aterm_i$ is an instance
atom. Moreover, assuming $\archtype = \sum_{i=1}^k \prod_{j=1}^{h_i}
P_{ij}(x_{ij})$, we denote $\Inter{\arule} \isdef
\{\{P_{ij}(x_{ij}) \mid j \in \interv{1}{h_i}\} \mid i \in
\interv{1}{k}\}$ the set of interactions occurring in $\arule$.

\begin{figure}[!t]
\caption{Definition of the $\mathit{Flow}$ Formula}
\label{fig:flow}
\vspace*{-.5\baselineskip}
\begin{eqnarray}
\mathit{Flow}(\vec{X}, \vec{Y}, \vec{U}) & \isdef & 
\bigvee_{1 \leq i \leq N} \bigvee_{\aint \in \Inter{\arule_i}} \mathit{IFlow}_{i,\aint}(\vec{X}, \vec{Y}, \vec{U}) \label{eq:flow} 
\end{eqnarray}
\vspace*{-\baselineskip}
\begin{eqnarray}
\mathit{IFlow}_{\ell,\set{P_1(x_1), \ldots, P_n(x_n)}}(\vec{X}, \vec{Y}, \vec{U}) \isdef \exists y_0 \ldots \exists y_n ~.~ U_\ell(y_0) ~\wedge \label{eq:iflow} \\
\bigwedge_{i=1}^n \Big(\bigvee_{
  \begin{array}{l}
    \scriptstyle{\arule' = \left(\apred'(x_1, \ldots, x_{\#(\apred')}) \leftarrow_{\asys_\abehterm} \abehterm'\right)} \\[-.5mm] 
    \scriptstyle{\abehtype(y_i) \in \instances{\abehterm'}}
\end{array}}
\mathit{Path}^{x_i,y_i}_{\arule_\ell,\arule'}(y_0,y_i,\vec{U})\Big) ~\wedge \nonumber \\
\forall x . \bigwedge_{S \in \bigcup_{j=1}^K \states_{\abehtype_j}} \Big[\Big(X_S(x) \leftrightarrow \bigvee_{\pre{P_k} = S} x = y_k\Big)
\wedge \Big(Y_S(x) \leftrightarrow \bigvee_{\post{P_k} = S} x = y_k\Big)\Big] \nonumber
\end{eqnarray}
\vspace*{-2\baselineskip}
\end{figure}
\begin{assumption}\label{ass:pre-post}
  For any component type $\abehtype = \tuple{\ports, \states,
    \initstate, \rules}$ and any two transition rules $S_1
  \arrow{P_1}{\abehtype} T_1, S_2 \arrow{P_2}{\abehtype} T_2$, if $P_1
  = P_2$ then $S_1 = S_2$ and $T_1 = T_2$. For a transition rule $S
  \arrow{P}{} T \in \rules_\abehtype$, let $\pre{P} \isdef S$ and
  $\post{P} \isdef T$ denote the pre- and post-state of the unique
  transition rule whose label is $P$.
\end{assumption}
The above assumption can be lifted at the cost of cluttering the
following presentation. The $\mathit{Flow}$ formula (\ref{eq:flow}) is
defined in Fig. \ref{fig:flow}. Essentially, $\mathit{Flow}$ is split
into a disjunction of $\mathit{IFlow}_{\ell,\set{P_1(x_1), \ldots,
    P_n(x_n)}}$ formul{\ae} (\ref{eq:iflow}), one for each set of
ports $\{P_1(x_1), \ldots, P_n(x_n)\}$ that denotes an interaction of
the rule $\arule_\ell$, for all $\ell \in \interv{1}{N}$. To
understand the formul{\ae} (\ref{eq:iflow}), recall that each of the
variables $x_1, \ldots, x_n$ is interpreted as the (unique) node of
the rewriting tree containing an instance atom $\abehtype_i(x_i)$. In
order to find this node, we track the variable $x_i$ from the current
node $y_0$, labeled by the rule $\arule_\ell$, to the node $y_i$,
where this instance atom occurs. This is done by the
$\mathit{Path}^{z,u}_{\arule,\arule'}(x,y,\vec{U})$ formula, that holds
iff $\mathcal{T} \in \rtrees{\abehterm}{\asys}$ is a rewriting tree,
uniquely encoded by the interpretation of the $\vec{U}$ variables, and
$x, y$ are mapped to the endpoints of a path from a node $w \in
\nodes(\mathcal{T})$, with label $\mathcal{T}(w) = \arule$ to a node
$w' \in \nodes(\mathcal{T})$, with label $\mathcal{T}(w') = \arule'$,
such that $z$ and $u$ are variables that occur in the bodies of
$\arule$ and $\arule'$, respectively, mapped to the same identifier
(node) in any ground term from the set
$\instant{\charterm{\mathcal{T}}{}}{}$. Note that, by the definition
of ground sets, two different variables are mapped to the same
identifier only if they are replaced by the same variable, when
$\charterm{\mathcal{T}}{}$ is built from the labels of $\mathcal{T}$
(Definition \ref{def:char-term}).

We encode sets of paths in a rewriting tree by a finite automaton and
use a classical result from automata theory to define
$\mathit{Path}^{z,u}_{\arule,\arule'}(x,y,\vec{U})$ by turning the
finite automaton into a \wss{\kappa}\ formula. But first, let us
define paths in a tree formally. Given a tree $\mathcal{T}$, with
$\nodes(\mathcal{T}) \subseteq \interv{0}{\kappa-1}^*$, a \emph{path}
is a finite sequence of nodes $\rho = n_1, \ldots, n_\ell$ such that,
for all $i \in \interv{1}{\ell-1}$, $n_{i+1}$ is either the parent
($n_i = n_{i+1} \alpha_i$) or a child ($n_{i+1} = n_i \alpha_i$) of
$n_i$, for some $\alpha_i \in \interv{0}{\kappa-1}$. The path is
determined by the source node and the sequence $(\alpha_1,d_1) \ldots
(\alpha_{\ell-1},d_{\ell-1})$ of \emph{directions} $(\alpha_i,d_i) \in
\interv{0}{\kappa-1} \times \set{\uparrow,\downarrow}$, with the
following meaning: $d_i = \uparrow$ if $n_{i+1} \alpha_i = n_i$ and
$d_i = \downarrow$ if $n_{i+1} = n_i \alpha_i$.

A \emph{path automaton} is a tuple $A = (Q,I,F,\delta)$, where $Q$ is
a set of states, $I, F \subseteq Q$ are the initial and final states,
respectively, and $\delta \subseteq Q \times \interv{0}{\kappa-1}
\times \set{\uparrow,\downarrow} \times Q$ is a set of transitions of
the form $q \arrow{(\alpha,d)}{} q'$, with $\alpha \in
\interv{0}{\kappa-1}$ being a direction and $d \in
\set{\uparrow,\downarrow}$ indicates whether the automaton moves up or
down in the tree. A run of $A$ over the path $\omega = (\alpha_1, d_1)
\ldots (\alpha_{n-1}, d_{n-1})$ is a sequence of states $q_1, \ldots,
q_n \in Q$ such that $q_1 \in I$ and $q_i \arrow{(\alpha_i, d_i)}{}
q_{i+1} \in \delta$, for all $i \in \interv{1}{n-1}$. The run is
accepting iff $q_n \in F$ and the \emph{language} of $A$ is the set of
paths over which $A$ has an accepting run, denoted $\lang{A}$.

A path automaton $A = (Q,I,F,\delta)$ corresponds, in the sense of
Lemma \ref{lemma:automata-wsks} below, to the following
\wss{\kappa}\ formula, that can be effectively built from the
description of $A$:
\[\begin{array}{l}
\Phi_A(x,y,\overline{\vec{X}}) \isdef 
\bigwedge_{1 \leq i \neq j \leq N} \forall z . \Big(\neg \overline{X}_i(z) \vee \neg \overline{X}_j(z)\Big)
\wedge \bigvee_{q_i \in I} \overline{X}_i(x) ~\wedge~ \bigvee_{q_j \in F} \overline{X}_j(y) ~\wedge \\[1mm]
\bigwedge_{i=1}^N \forall z ~.~ z \neq y \wedge \overline{X}_i(z) \rightarrow
\Big(\bigvee_{\scriptstyle{q_i \arrow{(\alpha,\downarrow)}{} q_j}} \overline{X}_j(\succ_\alpha(z))
\vee \bigvee_{\scriptstyle{q_i \arrow{(\alpha,\uparrow)}{} q_j}} \exists z' ~.~ \succ_\alpha(z') = z \wedge \overline{X}_j(z')\Big) \\[1mm]
\bigwedge_{i=1}^N \forall z ~.~ z \neq x \wedge \overline{X}_j(z) \rightarrow
\Big(\bigvee_{\scriptstyle{q_i \arrow{(\alpha,\downarrow)}{} q_j}} \exists z' ~.~ \succ_\alpha(z') = z \wedge \overline{X}_i(z')
\vee \bigvee_{\scriptstyle{q_i \arrow{(\alpha,\uparrow)}{} q_j}} \overline{X}_i(\succ_\alpha(z))\Big)
\end{array}\]
where $Q = \set{q_1, \ldots, q_L}$ and $\overline{\vec{X}} =
\tuple{\overline{X}_1, \ldots, \overline{X}_L}$ are second order
variables interpreted as the sets of tree nodes labeled by the
automaton with $q_1, \ldots, q_L$, respectively. Intuitively, the
first three conjuncts of the above formula encode the facts that
$\overline{\vec{X}}$ are disjoint (no tree node is labeled by more
than one state during the run), the run starts in an initial state
with node $x$ and ends in a final state with node $y$. The fourth
conjunct states that, for every non-final node on the path, if the
automaton visits that node by state $q_i$, then either the node has a
$\dn{\alpha}$-child or a $\up{\alpha}$-parent visited by state $q_j$,
where $q_i \arrow{\dn{\alpha}}{} q_j$ and $q_i \arrow{\up{\alpha}}{}
q_j$ are transitions of the automaton. The fifth conjunct is the
reversed flow condition on the path, needed to ensure that
$\overline{\vec{X}}$ do not contain useless nodes, being thus
symmetric to the fourth. The following lemma is adapted from folklore
automata-logic connection results\footnote{A similar conversion of
  tree walking automata to \mso\ has been described in
  \cite{DBLP:conf/cade/IosifRS13}.} \cite[\S2.10]{KhoussainovNerode}:
\begin{lemma}\label{lemma:automata-wsks}
  Given a tree $\mathcal{T}$ with $\nodes(\mathcal{T}) \subseteq
  \interv{0}{\kappa-1}^*$ and a path $\omega \in (\interv{0}{\kappa-1}
  \times \set{\uparrow, \downarrow})^*$ from $w_1$ to $w_2$ in
  $\mathcal{T}$, we have $\omega \in \lang{A}$ iff $[x \leftarrow w_1,
    y \leftarrow w_2] \models \exists \overline{\vec{X}} ~.~
  \Phi_A(x,y,\overline{\vec{X}})$.
\end{lemma}

\begin{figure}[t!]
\caption{Path Automata Recognizing the Instantiation Paths from Example
  \ref{ex:tll}}
\label{fig:tll-auto}
\vspace*{-.5\baselineskip}
\centerline{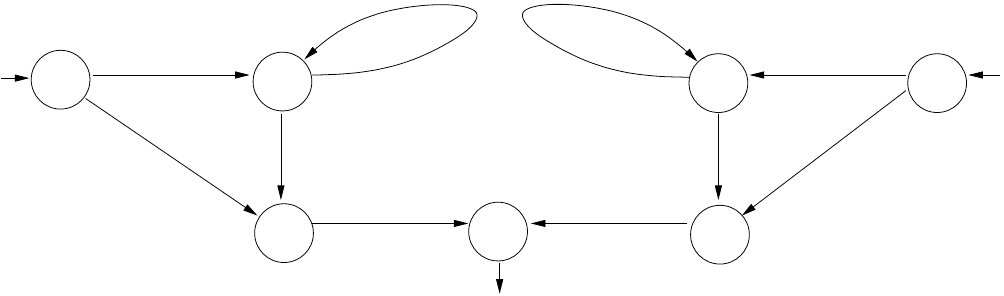}
\vspace*{-\baselineskip}
\end{figure}

Our purpose is to define path automata that recognize the paths
between the node where a bound variable is introduced and the node
where the variable is instantiated, in a given rewriting tree. For
example, the paths that track the instantiations of the variables
$l^\epsilon_1$ and $r^\epsilon_1$ in the rewriting tree for the term
$\mathit{Root}()$ generated by the rewriting system from Example
\ref{ex:tll} are depicted in red in Fig. \ref{fig:tll}. To this end,
we define a path automaton that tracks the instantiation of variables
from the rewriting system $\asys_\abehterm$. For each pair of rules
$\arule_1, \arule_2 \in \asys$ and variables $z_1,z_2 \in \vars$ that
occur in the bodies of $\arule_1$ and $\arule_2$, respectively, we
define $A^{z_1,z_2}_{\arule_1,\arule_2} \isdef
(Q,I^{z_1}_{\arule_1},F^{z_2}_{\arule_2},\delta)$ as follows.  We
associate a state $q^d_{\arule,z}$ to each rule $\arule = (\apred(x_1,
\ldots, x_{\#\apred}) \leftarrow_{\asys_\abehterm} \abehterm')$, each
variable $z$ occurring (free or bound) in $\abehterm'$ and each
direction $d \in \set{\uparrow,\downarrow}$. The sets of initial and
final states are $I^{z_1}_{\arule_1} \isdef \{q^d_{\arule_1,z_1} \mid
d = \uparrow,\downarrow\}$ and $F^{z_2}_{\arule_2} \isdef
\{q^\downarrow_{\arule_2,z_2}\}$. The transition relation consists of
the triples \(q^\downarrow_{\arule_1,y_j} \arrow{\dn{\alpha}}{}
q^\downarrow_{\arule_2,x_j}\), \(q^\uparrow_{\arule_2,x_j}
\arrow{\up{\alpha}}{} q^\uparrow_{\arule_1,y_j}\) and
\(q^\uparrow_{\arule_2,x_j} \arrow{\up{\alpha}}{}
q^\downarrow_{\arule_1,y_j}\), for any two distinct rules $\arule_i =
(\apred_j(x_1, \ldots, x_{\#(\apred)}) \leftarrow_{\asys_\abehterm}
\abehterm_i)$, $i = 1,2$, all $\alpha \in
\interv{0}{\npred{\abehterm_1}}$, such that
$\pred{\alpha}{\abehterm_1} = \apred_2(y_1,\ldots,y_{\#(\apred_2)})$
and all $j \in \interv{1}{\#(\apred_2)}$.  For instance, the path
automata that recognize the instantiation paths for the variables
$l^\epsilon_1$ and $r^\epsilon_1$ in the rewriting tree for the term
$\mathit{Root}()$ generated by the rewriting system from Example
\ref{ex:tll} are depicted in Fig. \ref{fig:tll-auto}. The initial
states are $q^\downarrow_{\text{\ref{rule:tll-root}},l_1}$ and
$q^\downarrow_{\text{\ref{rule:tll-root}},r_1}$, respectively, and the
final state is $q^\downarrow_{\text{\ref{rule:tll5}},n}$ in both
cases, where the labels of the rules of the rewriting system are the
ones from Example \ref{ex:tll}. We define the
$\mathit{Path}^{z_1,z_2}_{\arule_1,\arule_2}$ formula following the
below lemma, proving the correctness of the automata construction:
\begin{lemma}\label{lemma:path-automaton}
  Let $\mathcal{T} \in \rtrees{\abehterm}{\asys}$ be a rewriting tree
  and $w_i \in \nodes(\mathcal{T})$ be nodes labeled with the rules
  $\mathcal{T}(w_i) = \arule_i =
  \left(\apred_i(x_{i,1},\ldots,x_{i,\#(\apred_i)})
  \leftarrow_{\asys_\abehterm} \abehterm_i\right)$, for $i =
  1,2$. Then, for all $k_i \in \interv{1}{\#(\apred_i)}$, $i=1,2$, the
  following are equivalent: \begin{compactenum}
  \item\label{it1:path-automaton} $x_{1,k_1}$ and $x_{2,k_2}$ are
    mapped to the same identifier in any ground term $\aterm \in
    \instant{\charterm{\mathcal{T}}{}}{}$,
  \item\label{it2:path-automaton}
    $A_{\arule_1,\arule_2}^{x_{1,k_1}\!,~x_{2,k_2}}$ accepts the sequence
    of directions labeling the path from $w_1$ to $w_2$ in
    $\mathcal{T}$.
  \end{compactenum}
\end{lemma}
\proof{ Let $\omega \isdef \up{\alpha_1}, \ldots, \up{\alpha_i},
  \dn{\alpha_{i+1}}, \ldots, \dn{\alpha_n}$ be the sequence of
  directions labeling the path $\rho \isdef (w_1 = v_1, \ldots,
  v_{n+1} = w_2)$ and $\mathcal{T}(v_j) \isdef \overline{\arule}_j
  \isdef \left(\apred_{j}(x^j_1, \ldots, x^j_{\#(\apred_{j})})
  \leftarrow \abehterm_{j}\right)$, for all $j \in \interv{1}{n+1}$,
  where $\overline{\arule}_1 = \arule_1$ and $\overline{\arule}_{n+1}
  = \arule_2$.

  \noindent (\ref{it1:path-automaton}) ``$\Rightarrow$''
  (\ref{it2:path-automaton}) By the definition of ground sets,
  $x_{1,k_1}$ and $x_{2,k_2}$ are mapped to the same identifier in any
  ground term $\aterm \in \instant{\charterm{\mathcal{T}}{}}{}$ iff
  they are substituted by the same variable in
  $\charterm{\mathcal{T}}{}$, by the composed substitution along the
  path $\rho$. More precisely, let $x_{1,k_1} = z_1 \leftarrow \ldots
  \leftarrow z_i \rightarrow z_{i+1} \rightarrow \ldots \rightarrow
  z_{n+1} = x_{2,k_2}$ be the sequence of substitutions that match
  $x_{1,k_1}$ and $x_{2,k_2}$. By Definition \ref{def:rewriting-tree},
  we have: \begin{compactitem}
  \item for each $j \in \interv{1}{i}$,
    $\pred{\alpha_j}{\abehterm_{j+1}} = \apred_{j}(y_1, \ldots,
    y_{\#(\apred_{j})})$ and each $x^j_\ell$ is substituted by
    $y_\ell$, for all $\ell \in \interv{1}{\#(\apred_{j})}$. Then
    there exists $\ell \in \interv{1}{\#(\apred_{j})}$ such that the
    mapping $x^j_\ell = z_j \leftarrow z_{j+1} = y_\ell$ is induced by
    this substitution. Then, there exist transitions
    $q^\uparrow_{\overline{\arule}_j,z_j} \arrow{\up{\alpha_j}}{}
    q^\uparrow_{\overline{\arule}_{j+1},z_{j+1}}$, for all $j \in
    \interv{1}{i-1}$, and $q^\uparrow_{\overline{\arule}_i,z_i}
    \arrow{\up{\alpha_i}}{}
    q^\downarrow_{\overline{\arule}_{i+1},z_{i+1}}$ in $\delta$, by
    definition.
  \item for each $j \in \interv{i+1}{n}$,
    $\pred{\alpha_j}{\abehterm_{j}} = \apred_{j+1}(y_1, \ldots,
    y_{\#(\apred_{j+1})})$ and each $x^{j+1}_\ell$ is substituted by
    $y_\ell$, for all $\ell \in \interv{1}{\#(\apred_{j+1})}$. By a
    similar argument as above, there exists a transition
    $q^\downarrow_{\overline{\arule}_j,z_j} \arrow{\dn{\alpha_j}}{}
    q^\downarrow_{\overline{\arule}_{j+1},z_{j+1}}$ in $\delta$.
  \end{compactitem}
  By chaining the above transitions, we obtain a run of
  $A_{\arule_1,\arule_2}^{x_{1,k_1},x_{2,k_2}}$ over $\omega$. By the
  definitions of $I_{\arule_1}^{x_{1,k_1}}$ and
  $F_{\arule_2}^{x_{2,k_2}}$, this run is accepting, hence $\omega \in
  \lang{A_{\arule_1,\arule_2}^{x_{1,k_1},x_{2,k_2}}}$.

  \noindent(\ref{it2:path-automaton}) ``$\Rightarrow$''
  (\ref{it1:path-automaton}) Let $q^\uparrow_{\arule_1,x_{1,k_1}} =
  q^\uparrow_{\overline{\arule}_1,z_1} \arrow{\up{\alpha_1}}{} \ldots
  \arrow{\up{\alpha_i}}{}
  q^\downarrow_{\overline{\arule}_{i+1},z_{i+1}}
  \arrow{\dn{\alpha_{i+1}}}{} \ldots \arrow{\dn{\alpha_n}}{}
  q^\downarrow_{\overline{\arule}_{n+1},z_{n+1}} =
  q^\downarrow_{\arule_2,x_{2,k_2}}$ be an accepting run of
  $A_{\arule_1,\arule_2}^{x_{1,k_1},x_{2,k_2}}$ over $\omega$.  We
  give the proof only in the case the initial state on the run is
  labeled with $\uparrow$ and $i \in \interv{2}{n}$ is the position
  where the label changes to $\downarrow$. In the other case (the
  initial state is labeled with $\downarrow$) the entire path is
  labeled with $\downarrow$ and the argument is similar. By the
  definition of $A_{\arule_1,\arule_2}^{x_{1,k_1},x_{2,k_2}}$, there
  exists a sequence of substitutions $x_{1,k_1} = z_1 \leftarrow
  \ldots \leftarrow z_{i+1} \rightarrow \ldots \rightarrow z_{n+1} =
  x_{2,k_2}$, by which $x_{1,k_1}$ and $x_{2,k_2}$ are mapped to the
  same variable in $\charterm{\mathcal{T}}{}$. Hence $x_{1,k_1}$ and
  $x_{2,k_2}$ are always mapped to the same identifier in each ground
  term $\aterm \in \instant{\charterm{\mathcal{T}}{}}{}$. \qed}

\vspace*{-\baselineskip}
\[\begin{array}{rcl}
\mathit{Path}^{z_1,z_2}_{\arule_1,\arule_2}(x,y,\vec{U}) & \isdef & 
\exists \overline{X}_1 \ldots \exists \overline{X}_L ~.~
\Phi_{A^{z_1,z_2}_{\arule_1,\arule_2}}(x,y,\overline{\vec{X}}) \wedge
\Psi(\overline{\vec{X}}, \vec{U}) \\
\Psi(\overline{\vec{X}},\vec{U}) & \isdef & 
 \bigwedge_{d = \uparrow,\downarrow} \bigwedge_{\arule_i =
   \left(\apred'(x_1, \ldots, x_{\#(\apred')})
   \leftarrow_{\asys_\abehterm} \abehterm'\right)} \bigwedge_{z \in
   \fv{\abehterm'}} \forall x ~.~ \overline{X}^d_{\arule,z}(x)
 \rightarrow U_i(x)
\end{array}\]
The formula $\Psi$ states that all nodes labeled with a state
$q^d_{\arule,z}$ during the run must be also labeled with $\arule$ in
the rewriting tree. The lemma below proves that the definition
(\ref{eq:flow}) of the formula $\mathit{Flow}$ meets condition
(\ref{eq:flow-cond}):

\begin{lemma}\label{lemma:flow}
  For any valuation $\nu : \vec{X} \cup \vec{Y} \cup \vec{U} \cup
  \vec{Z} \rightarrow 2^{\ids}$, such that $\nu \models
  \mathit{RTree}(\vec{U}) \wedge \mathit{Inst}(\vec{U}, \vec{Z})
  \wedge \mathit{Config}(\vec{X}, \vec{Z}) \wedge
  \mathit{Config}(\vec{Y}, \vec{Z})$, the following are
  equivalent: \begin{compactenum}
  \item\label{it1:flow} $\nu \models \mathit{Flow}(\vec{X}, \vec{Y},
    \vec{U})$,
  \item\label{it2:flow} $\nu(\vec{X}) \rhd \pre{\pi}$ and
    $\nu(\vec{Y}) \rhd \post{\pi}$, for some interaction $\pi \in
    \aarch(\nu(\vec{U}))$.
  \end{compactenum}
\end{lemma}
\proof{ Since $\nu \models \mathit{RTree}(\vec{U})$, the tuple of sets
  $\nu(\vec{U})$ is parameter-compatible with $\asys$ and $\abehterm$,
  thus $\aarch(\nu(\vec{U}))$ is defined. Moreover, because $\nu
  \models \mathit{Inst}(\vec{U}, \vec{Z}) \wedge
  \mathit{Config}(\vec{X}, \vec{Z}) \wedge \mathit{Config}(\vec{Y},
  \vec{Z})$, we have that $\nu(\vec{X})$ and $\nu(\vec{Y})$ denote
  valid configurations of $\mathcal{C}$.

  \noindent ``(\ref{it1:flow}) $\Rightarrow$ (\ref{it2:flow})'' Let
  $\mathcal{T} \in \rtrees{\abehterm}{\asys}$ be the rewriting tree
  uniquely determined by $\nu(\vec{U})$. Because $\nu \models
  \mathit{Flow}_{\asys,\abehterm}(\vec{X}, \vec{Y}, \vec{U})$, there
  exists a rule $\arule_\ell \in \asys_\abehterm$ and a set of terms
  $\set{P_1(x_1), \ldots P_n(x_n)} \in \Inter{\arule}$, such that $\nu
  \models \mathit{IFlow}_{\ell, \set{P_1(x_1), \ldots
      P_n(x_n)}}(\vec{X}, \vec{Y}, \vec{U})$ (\ref{eq:flow}). Let
  $w_0, \ldots, w_n \in \interv{0}{\kappa-1}^*$ be nodes, such that,
  by (\ref{eq:iflow}), \(\nu[y_0 \leftarrow w_0]
  \models U_\ell(y_0)\) and, for all $i \in \interv{1}{n}$:
  \[\begin{array}{rcl}
  \nu[y_i \leftarrow w_i] & \models & \mathit{Path}_{\arule,x_i,\arule',y_i}(y_0,y_i,\vec{U}) \\  
  && \text{for some } \arule' = \left(\apred'(x_1, \ldots, x_{\#(\apred')}) \leftarrow_{\asys_\abehterm} \abehterm'\right) \\
  && \text{ and } \abehtype(y_i) \in \instances{\abehterm'} \\  
  \nu[y_i \leftarrow w_i] & \models & \forall x . \Big(X_S(x) \leftrightarrow \bigvee_{\pre{P_i} = S} x = y_i\Big) 
  \text{, for all } S \in \bigcup_{j=1}^K \states_{\abehtype_j} \\
  \nu[y_i \leftarrow w_i] & \models & \forall x . \Big(Y_S(x) \leftrightarrow \bigvee_{\post{P_i} = S} x = y_i\Big) 
  \text{, for all } S \in \bigcup_{j=1}^K \states_{\abehtype_j}
  \end{array}\]
  Then $w_0 \in \nodes(\mathcal{T})$ and, since each variable $y_i$ is
  instantiated exactly once in $\charterm{\mathcal{T}}{}$, being
  assigned to $w_i$ (Assumption \ref{ass:one-inst}) and, by Lemma
  \ref{lemma:path-automaton}, there is a unique path in $\mathcal{T}$
  between $w_0$ and $w_i \in \nodes(\mathcal{T})$, for all $i \in
  \interv{1}{n}$. Then $\set{P_1(w_1), \ldots, P_n(w_n)} \in
  \aarch(\nu(\vec{U}))$ is the interaction defined by the rule $\arule
  = \mathcal{T}(w_0)$. Moreover, each set $\nu(X_S)$
  (resp. $\nu(Y_S)$), for $S \in \bigcup_{j=1}^K
  \states_{\abehtype_j}$, consists of the identifiers of those
  instances that are in the state $\pre{P_i(w_i)}$
  (resp. $\post{P_i(w_i)}$).

  \noindent''(\ref{it2:flow}) $\Rightarrow$ (\ref{it1:flow})'' By the
  definition of $\aarch$, each interaction $\pi \in
  \aarch(\nu(\vec{U}))$ corresponds to a node $w_0 \in
  \nodes(\mathcal{T})$ of a rewriting tree $\mathcal{T} \in
  \rtrees{\abehterm}{\asys}$, labeled by a rule $\arule_\ell =
  \mathcal{T}(w_0)$. Then let $\set{P_1(y_1), \ldots, P_n(y_n)}$ be a
  set of terms and $w_1, \ldots, w_n \in \nodes(\mathcal{T})$ be
  nodes, such that $\pi = \set{P_1(w_1), \ldots, P_n(w_n)}$. Since
  each variable $y_i$ is instantiated exactly once in
  $\charterm{\mathcal{T}}{}$ (Assumption \ref{ass:one-inst}), there
  exists a unique path from $w_0$ to $w_i$ in $\mathcal{T}$, for all
  $i \in \interv{1}{n}$. It is easy to check that $\nu \models
  \mathit{IFlow}_{\ell, \set{P_1(y_1), \ldots, P_n(y_n)}}(\vec{X},
  \vec{Y}, \vec{U})$, thus $\nu \models \mathit{Flow}_{\asys,
    \abehterm}(\vec{X}, \vec{Y}, \vec{U})$. \qed}

Together with Lemma \ref{lemma:trap-invariant}, this ensures that the
trap invariant of the parametric system corresponding to $\asys$ and
$\abehterm$ is defined in \wss{\kappa}, by the $\mathit{TrapInv}$
formula (\ref{eq:trap-inv}). Hence the verification of safety
properties (such as absence of deadlocks) is reduced to checking the
satisfiability of the $\mathit{Safe}$ formula (\ref{eq:safety}),
leading to the following result:

\begin{theorem}\label{thm:parametric-safety}
  Given a closed behavioral term $\abehterm$, a rewriting system
  $\asys$, a formula $\mathit{Bad}(\vec{X},\vec{U})$ and a tuple of
  sets $T_1, \ldots, T_N \subseteq \ids$, that are
  parameter-compatible with $\asys$ and $\abehterm$, the behavior
  $\mathcal{C}(\vec{T})$ is safe w.r.t the set of
  configurations $\errstates \isdef \{\sigma \mid \nu[U_1 \leftarrow
    T_1, \ldots, U_N \leftarrow T_N](\vec{X}) \rhd \sigma,~ \nu[U_1
    \leftarrow T_1, \ldots, U_N \leftarrow T_N] \models
  \mathit{Bad}(\vec{X},\vec{U})\}$ if $\mathit{Safe}(\vec{U})$ is
  unsatisfiable.
\end{theorem}
\proof{ We prove the contrapositive statement. Let $T_1, \ldots, T_N
  \subseteq \ids$ be sets such that $[U_1 \leftarrow T_1, \ldots, U_N
    \leftarrow T_N] \models \mathit{RTree}(\vec{U})$. Since
  $\mathcal{C}(\vec{T})$ is unsafe
  w.r.t. $\errstates$ and
  $\Theta(\mathcal{C}(\vec{T}))$ is an invariant for
  $\mathcal{C}(\vec{T})$, there exists a
  configuration $\sigma \in
  \Theta(\mathcal{C}(\vec{T})) \cap \errstates$. Let
  $\nu$ be a valuation such that $\nu(\vec{X}) \rhd \sigma$ and
  $\nu(U_i) = T_i$, for all $i \in \interv{1}{N}$. We prove that $\nu$
  is a model of $\mathit{RTree}(\vec{U})$, $\mathit{TrapInv}(\vec{X},
  \vec{U})$ and $\mathit{Bad}(\vec{X}, \vec{U})$, which suffices to
  prove that $\nu \models \mathit{Safe}(\vec{U})$, by
  (\ref{eq:safety}). Clearly, $\nu \models \mathit{RTree}(\vec{U})$
  because $[U_1 \leftarrow T_1, \ldots, U_N \leftarrow T_N] \models
  \mathit{RTree}(\vec{U})$ and $\nu \models \mathit{Bad}(\vec{X},
  \vec{U})$, because $\sigma \in \errstates$, by the definition of
  $\errstates$. Moreover, since the definition of
  $\mathit{Flow}(\vec{X}, \vec{Y}, \vec{U})$ meets condition
  (\ref{eq:flow}), by Lemma \ref{lemma:flow}, we obtain that $\nu
  \models \mathit{TrapInv}(\vec{X}, \vec{U})$, by Lemma
  \ref{lemma:trap-invariant}. This concludes our proof. \qed}

\vspace*{-\baselineskip}
\section{Experimental Evaluation}

We implemented the trap invariant synthesis in a prototype
tool\footnote{Available online at \url{https://github.com/raduiosif/rtab}.}
that generates the \wss{\kappa} formula corresponding to the
(sufficient) deadlock freedom condition (\ref{eq:safety}) from a given
behavioral term and a rewriting system. Our test cases are
hand-crafted examples of common architectures encountered in practice
(e.g.\ pipelines and stars), textbook examples (dining philosophers)
and several hierarchical tree-shaped architectures with rather complex
architectural patterns (trees with root links or leaves linked in a
ring).

The table below shows the results of checking deadlock freedom of
several test cases. The 2nd column gives the number of states $n_1
\times \ldots \times n_K$, where $n_i$ is the number of states in the
$i$-th component type and $K$ is the number of component types from
the system. The number of rewriting rules and interactions in the
specification are given in the 3rd and 4th columns, respectively. The
5th column reports the result of the satisfiability check
(\ref{eq:safety}) using the \textsc{Mona} v1.4-18 tool \cite{Mona} and
the 6th column shows the runing times (in seconds) on an Debian AMD64
2GHz machine with 16GB of RAM. The 7th and 8th columns report the type
of invariant (trap or $1$-invariant) used to prove deadlock freedom
and the 9th column gives the type of \wss{\kappa} logic, for $\kappa
\in \{1,2\}$.

\enlargethispage{2cm}
\begin{table}[h!]
\vspace*{-\baselineskip}
{\footnotesize\begin{center}
  \input{experiments}

\end{center}}
\vspace*{-2\baselineskip}
\end{table} 

The {\sf ring}, {\sf star} and {\sf ring-star} test cases correspond
to a simple token-ring, a star with one master (coordinator) and $n
\geq 2$ slaves and a star with $n$ slaves linked in a token-ring.

The {\sf alt-philo-sym} and {\sf alt-philo-asym} examples correspond
to the dining philosophers in which the philosophers pick their left
and right forks separately, with all symmetric philosophers and one
asymetric philosopher, respectively. The {\sf sync-philo} example
models the dining philosophers in which every philosopher picks her
forks simultaneously. It is known that {\sf alt-philo-sym} reaches a
deadlock configuration, whereas {\sf alt-philo-asym} and {\sf
  sync-philo} are deadlock free. Moreover, the {\sf alt-philo-asym}
system cannot be the proved deadlock free using trap invariants only
\cite[Proposition 1]{DBLP:conf/tacas/BozgaEISW20}. Following the
solution from \cite{DBLP:conf/tacas/BozgaEISW20}, we used the
structural information given by the $\mathit{Flow}$ formula
(\ref{eq:flow}) to synthethize \emph{$1$-invariants}, i.e.\ inductive
sets of configurations that contain exactly one active state at the
time\footnote{We refer the reader to \cite[Definition
    1]{DBLP:conf/tacas/BozgaEISW20} for a formal definition of
  $1$-invariants.}.

The {\sf tree-dfs} example models a binary tree architecture traversed
by a token in depth-first order, while the\begin{inparaenum}[(i)]
\item {\sf tree-back-root} and
\item {\sf tree-linked-leaves} (Example \ref{ex:tll}) 
\end{inparaenum}
go beyond trees, modeling hierarchical systems with parent-children
communication on top of which\begin{inparaenum}[(i)]
\item the nodes communicate with the root and 
\item the leaves are linked in a token-ring,
\end{inparaenum} 
respectively. 

\vspace*{-.5\baselineskip}
\section{Conclusions and Future Work}

We present a formal language for the specification of distributed
systems parameterized by the number of replicated components and by
the shape of the coordinating architecture. The language uses
inductive definitions to describe systems of unbounded size. We propose
a verification method for safety properties based on the synthesis of
structural invariants able to prove deadlock freedom for a number of
non-trivial models. 

One of the drawbacks that prevented us from tackling more real-life
examples is the lack of support for broadcast communication
(i.e.\ interactions that involve an unbounded number of
participants). We plan on adding support for broadcast in our
behavioral term algebra and develop further the invariant synthesis
method to take broadcast into account, as future work.

%% file: ring.pdf_t
\begin{picture}(0,0)%
\includegraphics{ring.pdf}%
\end{picture}%
\setlength{\unitlength}{1973sp}%
\begingroup\makeatletter\ifx\SetFigFont\undefined%
\gdef\SetFigFont#1#2#3#4#5{%
  \reset@font\fontsize{#1}{#2pt}%
  \fontfamily{#3}\fontseries{#4}\fontshape{#5}%
  \selectfont}%
\fi\endgroup%
\begin{picture}(12999,4660)(3814,-809)
\put(13126,2639){\makebox(0,0)[b]{\smash{{\SetFigFont{6}{7.2}{\rmdefault}{\mddefault}{\updefault}{\color[rgb]{1,0,0}\ref{rule:chain2}}%
}}}}
\put(5926,2639){\makebox(0,0)[b]{\smash{{\SetFigFont{5}{6.0}{\rmdefault}{\mddefault}{\updefault}{\color[rgb]{0,0,0}$\mathit{in}$}%
}}}}
\put(6151,2264){\makebox(0,0)[b]{\smash{{\SetFigFont{5}{6.0}{\rmdefault}{\mddefault}{\updefault}{\color[rgb]{0,0,0}$\mathit{out}$}%
}}}}
\put(6076,1964){\makebox(0,0)[b]{\smash{{\SetFigFont{5}{6.0}{\rmdefault}{\mddefault}{\updefault}{\color[rgb]{0,0,0}$q_1$}%
}}}}
\put(6076,3014){\makebox(0,0)[b]{\smash{{\SetFigFont{5}{6.0}{\rmdefault}{\mddefault}{\updefault}{\color[rgb]{0,0,0}$q_0$}%
}}}}
\put(7651,2639){\makebox(0,0)[b]{\smash{{\SetFigFont{5}{6.0}{\rmdefault}{\mddefault}{\updefault}{\color[rgb]{0,0,0}$\mathit{in}$}%
}}}}
\put(7876,2264){\makebox(0,0)[b]{\smash{{\SetFigFont{5}{6.0}{\rmdefault}{\mddefault}{\updefault}{\color[rgb]{0,0,0}$\mathit{out}$}%
}}}}
\put(7801,1964){\makebox(0,0)[b]{\smash{{\SetFigFont{5}{6.0}{\rmdefault}{\mddefault}{\updefault}{\color[rgb]{0,0,0}$q_1$}%
}}}}
\put(7801,3014){\makebox(0,0)[b]{\smash{{\SetFigFont{5}{6.0}{\rmdefault}{\mddefault}{\updefault}{\color[rgb]{0,0,0}$q_0$}%
}}}}
\put(9151,2639){\makebox(0,0)[b]{\smash{{\SetFigFont{5}{6.0}{\rmdefault}{\mddefault}{\updefault}{\color[rgb]{0,0,0}$\mathit{in}$}%
}}}}
\put(9376,2264){\makebox(0,0)[b]{\smash{{\SetFigFont{5}{6.0}{\rmdefault}{\mddefault}{\updefault}{\color[rgb]{0,0,0}$\mathit{out}$}%
}}}}
\put(9301,1964){\makebox(0,0)[b]{\smash{{\SetFigFont{5}{6.0}{\rmdefault}{\mddefault}{\updefault}{\color[rgb]{0,0,0}$q_1$}%
}}}}
\put(9301,3014){\makebox(0,0)[b]{\smash{{\SetFigFont{5}{6.0}{\rmdefault}{\mddefault}{\updefault}{\color[rgb]{0,0,0}$q_0$}%
}}}}
\put(4351,2639){\makebox(0,0)[b]{\smash{{\SetFigFont{5}{6.0}{\rmdefault}{\mddefault}{\updefault}{\color[rgb]{0,0,0}$\mathit{in}$}%
}}}}
\put(4576,2264){\makebox(0,0)[b]{\smash{{\SetFigFont{5}{6.0}{\rmdefault}{\mddefault}{\updefault}{\color[rgb]{0,0,0}$\mathit{out}$}%
}}}}
\put(4501,1964){\makebox(0,0)[b]{\smash{{\SetFigFont{5}{6.0}{\rmdefault}{\mddefault}{\updefault}{\color[rgb]{0,0,0}$q_1$}%
}}}}
\put(4501,3014){\makebox(0,0)[b]{\smash{{\SetFigFont{5}{6.0}{\rmdefault}{\mddefault}{\updefault}{\color[rgb]{0,0,0}$q_0$}%
}}}}
\put(6901,3014){\makebox(0,0)[b]{\smash{{\SetFigFont{6}{7.2}{\rmdefault}{\mddefault}{\updefault}{\color[rgb]{0,0,0}$\cdots$}%
}}}}
\put(5101,3239){\makebox(0,0)[b]{\smash{{\SetFigFont{5}{6.0}{\rmdefault}{\mddefault}{\updefault}{\color[rgb]{0,0,0}$\mathit{out}$}%
}}}}
\put(6676,3239){\makebox(0,0)[b]{\smash{{\SetFigFont{5}{6.0}{\rmdefault}{\mddefault}{\updefault}{\color[rgb]{0,0,0}$\mathit{out}$}%
}}}}
\put(8401,3239){\makebox(0,0)[b]{\smash{{\SetFigFont{5}{6.0}{\rmdefault}{\mddefault}{\updefault}{\color[rgb]{0,0,0}$\mathit{out}$}%
}}}}
\put(7201,2864){\makebox(0,0)[b]{\smash{{\SetFigFont{5}{6.0}{\rmdefault}{\mddefault}{\updefault}{\color[rgb]{0,0,0}$\mathit{in}$}%
}}}}
\put(3901,2864){\makebox(0,0)[b]{\smash{{\SetFigFont{5}{6.0}{\rmdefault}{\mddefault}{\updefault}{\color[rgb]{0,0,0}$\mathit{in}$}%
}}}}
\put(5476,2864){\makebox(0,0)[b]{\smash{{\SetFigFont{5}{6.0}{\rmdefault}{\mddefault}{\updefault}{\color[rgb]{0,0,0}$\mathit{in}$}%
}}}}
\put(4576,3464){\makebox(0,0)[b]{\smash{{\SetFigFont{6}{7.2}{\rmdefault}{\mddefault}{\updefault}{\color[rgb]{0,0,0}$\mathit{CType}(y_1)$}%
}}}}
\put(6076,3464){\makebox(0,0)[b]{\smash{{\SetFigFont{6}{7.2}{\rmdefault}{\mddefault}{\updefault}{\color[rgb]{0,0,0}$\mathit{CType}(y^1_1)$}%
}}}}
\put(7801,3464){\makebox(0,0)[b]{\smash{{\SetFigFont{6}{7.2}{\rmdefault}{\mddefault}{\updefault}{\color[rgb]{0,0,0}$\mathit{CType}(y^n_1)$}%
}}}}
\put(9301,3464){\makebox(0,0)[b]{\smash{{\SetFigFont{6}{7.2}{\rmdefault}{\mddefault}{\updefault}{\color[rgb]{0,0,0}$\mathit{CType}(y_2)$}%
}}}}
\put(8701,2864){\makebox(0,0)[b]{\smash{{\SetFigFont{5}{6.0}{\rmdefault}{\mddefault}{\updefault}{\color[rgb]{0,0,0}$\mathit{in}$}%
}}}}
\put(9976,2864){\makebox(0,0)[b]{\smash{{\SetFigFont{5}{6.0}{\rmdefault}{\mddefault}{\updefault}{\color[rgb]{0,0,0}$\mathit{out}$}%
}}}}
\put(6901,1289){\makebox(0,0)[b]{\smash{{\SetFigFont{6}{7.2}{\rmdefault}{\mddefault}{\updefault}{\color[rgb]{0,0,0}(a)}%
}}}}
\put(8551,839){\makebox(0,0)[b]{\smash{{\SetFigFont{6}{7.2}{\rmdefault}{\mddefault}{\updefault}{\color[rgb]{0,0,0}$q_1$}%
}}}}
\put(8551,464){\makebox(0,0)[b]{\smash{{\SetFigFont{6}{7.2}{\rmdefault}{\mddefault}{\updefault}{\color[rgb]{0,0,0}$q_0$}%
}}}}
\put(8551, 14){\makebox(0,0)[b]{\smash{{\SetFigFont{6}{7.2}{\rmdefault}{\mddefault}{\updefault}{\color[rgb]{0,0,0}$q_0$}%
}}}}
\put(10651,839){\makebox(0,0)[b]{\smash{{\SetFigFont{6}{7.2}{\rmdefault}{\mddefault}{\updefault}{\color[rgb]{0,0,0}$q_0$}%
}}}}
\put(10651,464){\makebox(0,0)[b]{\smash{{\SetFigFont{6}{7.2}{\rmdefault}{\mddefault}{\updefault}{\color[rgb]{0,0,0}$q_1$}%
}}}}
\put(10651, 14){\makebox(0,0)[b]{\smash{{\SetFigFont{6}{7.2}{\rmdefault}{\mddefault}{\updefault}{\color[rgb]{0,0,0}$q_0$}%
}}}}
\put(12751,839){\makebox(0,0)[b]{\smash{{\SetFigFont{6}{7.2}{\rmdefault}{\mddefault}{\updefault}{\color[rgb]{0,0,0}$q_0$}%
}}}}
\put(12751,464){\makebox(0,0)[b]{\smash{{\SetFigFont{6}{7.2}{\rmdefault}{\mddefault}{\updefault}{\color[rgb]{0,0,0}$q_0$}%
}}}}
\put(12751, 14){\makebox(0,0)[b]{\smash{{\SetFigFont{6}{7.2}{\rmdefault}{\mddefault}{\updefault}{\color[rgb]{0,0,0}$q_1$}%
}}}}
\put(9601,539){\makebox(0,0)[b]{\smash{{\SetFigFont{6}{7.2}{\rmdefault}{\mddefault}{\updefault}{\color[rgb]{0,0,0}$\{\mathit{out}(y_1), \mathit{in}(y^1_1)\}$}%
}}}}
\put(11701,539){\makebox(0,0)[b]{\smash{{\SetFigFont{6}{7.2}{\rmdefault}{\mddefault}{\updefault}{\color[rgb]{0,0,0}$\{\mathit{out}(y^1_1), \mathit{in}(y_2)\}$}%
}}}}
\put(10651,-361){\makebox(0,0)[b]{\smash{{\SetFigFont{6}{7.2}{\rmdefault}{\mddefault}{\updefault}{\color[rgb]{0,0,0}$\{\mathit{in}(y_1), \mathit{out}(y_2)\}$}%
}}}}
\put(10651,-736){\makebox(0,0)[b]{\smash{{\SetFigFont{6}{7.2}{\rmdefault}{\mddefault}{\updefault}{\color[rgb]{0,0,0}(c)}%
}}}}
\put(13951,1514){\makebox(0,0)[b]{\smash{{\SetFigFont{6}{7.2}{\rmdefault}{\mddefault}{\updefault}{\color[rgb]{0,0,0}(b)}%
}}}}
\put(11551,2114){\makebox(0,0)[b]{\smash{{\SetFigFont{6}{7.2}{\rmdefault}{\mddefault}{\updefault}{\color[rgb]{0,0,0}$\mathit{CType}(y_1)$}%
}}}}
\put(13351,2114){\makebox(0,0)[b]{\smash{{\SetFigFont{6}{7.2}{\rmdefault}{\mddefault}{\updefault}{\color[rgb]{0,0,0}$\mathit{CType}(y^1_1)$}%
}}}}
\put(16426,2039){\makebox(0,0)[b]{\smash{{\SetFigFont{5}{6.0}{\rmdefault}{\mddefault}{\updefault}{\color[rgb]{0,0,0}$\mathit{out}$}%
}}}}
\put(15376,2114){\makebox(0,0)[b]{\smash{{\SetFigFont{6}{7.2}{\rmdefault}{\mddefault}{\updefault}{\color[rgb]{0,0,.82}$\mathit{Chain}(y^2_1,y_2)$}%
}}}}
\put(10951,1964){\makebox(0,0)[b]{\smash{{\SetFigFont{5}{6.0}{\rmdefault}{\mddefault}{\updefault}{\color[rgb]{0,0,0}$\mathit{in}$}%
}}}}
\put(12751,1964){\makebox(0,0)[b]{\smash{{\SetFigFont{5}{6.0}{\rmdefault}{\mddefault}{\updefault}{\color[rgb]{0,0,0}$\mathit{in}$}%
}}}}
\put(14551,1964){\makebox(0,0)[b]{\smash{{\SetFigFont{5}{6.0}{\rmdefault}{\mddefault}{\updefault}{\color[rgb]{0,0,0}$\mathit{in}$}%
}}}}
\put(12151,2264){\makebox(0,0)[b]{\smash{{\SetFigFont{5}{6.0}{\rmdefault}{\mddefault}{\updefault}{\color[rgb]{0,0,0}$\mathit{out}$}%
}}}}
\put(13951,2264){\makebox(0,0)[b]{\smash{{\SetFigFont{5}{6.0}{\rmdefault}{\mddefault}{\updefault}{\color[rgb]{0,0,0}$\mathit{out}$}%
}}}}
\put(16426,3614){\makebox(0,0)[b]{\smash{{\SetFigFont{6}{7.2}{\rmdefault}{\mddefault}{\updefault}{\color[rgb]{0,0,.82}$\mathit{Ring}()$}%
}}}}
\put(15376,2564){\makebox(0,0)[b]{\smash{{\SetFigFont{6}{7.2}{\rmdefault}{\mddefault}{\updefault}{\color[rgb]{0,0,.82}$\mathit{Chain}(y^1_1,y_2)$}%
}}}}
\put(15376,3014){\makebox(0,0)[b]{\smash{{\SetFigFont{6}{7.2}{\rmdefault}{\mddefault}{\updefault}{\color[rgb]{0,0,.82}$\mathit{Chain}(y_1,y_2)$}%
}}}}
\put(11326,3014){\makebox(0,0)[b]{\smash{{\SetFigFont{6}{7.2}{\rmdefault}{\mddefault}{\updefault}{\color[rgb]{1,0,0}\ref{rule:chain2}}%
}}}}
\put(10576,3614){\makebox(0,0)[b]{\smash{{\SetFigFont{6}{7.2}{\rmdefault}{\mddefault}{\updefault}{\color[rgb]{1,0,0}\ref{rule:ring}}%
}}}}
\end{picture}%

%% file: behavioral-types.pdf_t
\begin{picture}(0,0)%
\includegraphics{behavioral-types.pdf}%
\end{picture}%
\setlength{\unitlength}{2368sp}%
\begingroup\makeatletter\ifx\SetFigFont\undefined%
\gdef\SetFigFont#1#2#3#4#5{%
  \reset@font\fontsize{#1}{#2pt}%
  \fontfamily{#3}\fontseries{#4}\fontshape{#5}%
  \selectfont}%
\fi\endgroup%
\begin{picture}(8430,2431)(136,-1859)
\put(8551,-961){\makebox(0,0)[b]{\smash{{\SetFigFont{7}{8.4}{\rmdefault}{\mddefault}{\updefault}{\color[rgb]{0,0,0}$\set{S_1(i),T_1(j),T_2(k)}$}%
}}}}
\put(676,389){\makebox(0,0)[lb]{\smash{{\SetFigFont{7}{8.4}{\rmdefault}{\mddefault}{\updefault}{\color[rgb]{0,0,0}$S_1$}%
}}}}
\put(901,-1261){\makebox(0,0)[b]{\smash{{\SetFigFont{7}{8.4}{\rmdefault}{\mddefault}{\updefault}{\color[rgb]{0,0,0}$\mathit{Task}$}%
}}}}
\put(1201,-511){\makebox(0,0)[lb]{\smash{{\SetFigFont{7}{8.4}{\rmdefault}{\mddefault}{\updefault}{\color[rgb]{0,0,0}$\mathit{acq}$}%
}}}}
\put(151, 14){\makebox(0,0)[lb]{\smash{{\SetFigFont{7}{8.4}{\rmdefault}{\mddefault}{\updefault}{\color[rgb]{0,0,0}$\mathit{rel}$}%
}}}}
\put(2476,-1261){\makebox(0,0)[b]{\smash{{\SetFigFont{7}{8.4}{\rmdefault}{\mddefault}{\updefault}{\color[rgb]{0,0,0}$\mathit{Lock}$}%
}}}}
\put(2776,-511){\makebox(0,0)[lb]{\smash{{\SetFigFont{7}{8.4}{\rmdefault}{\mddefault}{\updefault}{\color[rgb]{0,0,0}$\mathit{lock}$}%
}}}}
\put(1426,-61){\makebox(0,0)[lb]{\smash{{\SetFigFont{7}{8.4}{\rmdefault}{\mddefault}{\updefault}{\color[rgb]{0,0,0}$\mathit{unlock}$}%
}}}}
\put(1651,-1786){\makebox(0,0)[lb]{\smash{{\SetFigFont{7}{8.4}{\rmdefault}{\mddefault}{\updefault}{\color[rgb]{0,0,0}(a)}%
}}}}
\put(2251,389){\makebox(0,0)[lb]{\smash{{\SetFigFont{7}{8.4}{\rmdefault}{\mddefault}{\updefault}{\color[rgb]{0,0,0}$S_2$}%
}}}}
\put(2251,-886){\makebox(0,0)[lb]{\smash{{\SetFigFont{7}{8.4}{\rmdefault}{\mddefault}{\updefault}{\color[rgb]{0,0,0}$T_2$}%
}}}}
\put(676,-886){\makebox(0,0)[lb]{\smash{{\SetFigFont{7}{8.4}{\rmdefault}{\mddefault}{\updefault}{\color[rgb]{0,0,0}$T_1$}%
}}}}
\put(5551,-61){\rotatebox{45.0}{\makebox(0,0)[b]{\smash{{\SetFigFont{7}{8.4}{\rmdefault}{\mddefault}{\updefault}{\color[rgb]{0,0,0}$\mathit{acq}(i)\cdot\mathit{lock}(k)$}%
}}}}}
\put(8101,-136){\rotatebox{315.0}{\makebox(0,0)[b]{\smash{{\SetFigFont{7}{8.4}{\rmdefault}{\mddefault}{\updefault}{\color[rgb]{0,0,0}$\mathit{acq}(j)\cdot\mathit{lock}(k)$}%
}}}}}
\put(6301,-661){\rotatebox{45.0}{\makebox(0,0)[b]{\smash{{\SetFigFont{7}{8.4}{\rmdefault}{\mddefault}{\updefault}{\color[rgb]{0,0,0}$\mathit{rel}(i)\cdot\mathit{unlock}(k)$}%
}}}}}
\put(7276,-736){\rotatebox{315.0}{\makebox(0,0)[b]{\smash{{\SetFigFont{7}{8.4}{\rmdefault}{\mddefault}{\updefault}{\color[rgb]{0,0,0}$\mathit{rel}(j)\cdot\mathit{unlock}(k)$}%
}}}}}
\put(6826,389){\makebox(0,0)[b]{\smash{{\SetFigFont{7}{8.4}{\rmdefault}{\mddefault}{\updefault}{\color[rgb]{0,0,0}$\set{S_1,(i),S_1(j),S_2(k)}$}%
}}}}
\put(4951,-961){\makebox(0,0)[b]{\smash{{\SetFigFont{7}{8.4}{\rmdefault}{\mddefault}{\updefault}{\color[rgb]{0,0,0}$\set{T_1(i),S_1(j),T_2(k)}$}%
}}}}
\end{picture}%

%% file: tll.pdf_t
\begin{picture}(0,0)%
\includegraphics{tll.pdf}%
\end{picture}%
\setlength{\unitlength}{1973sp}%
\begingroup\makeatletter\ifx\SetFigFont\undefined%
\gdef\SetFigFont#1#2#3#4#5{%
  \reset@font\fontsize{#1}{#2pt}%
  \fontfamily{#3}\fontseries{#4}\fontshape{#5}%
  \selectfont}%
\fi\endgroup%
\begin{picture}(12624,4449)(-911,-4573)
\put(8026,-2686){\makebox(0,0)[b]{\smash{{\SetFigFont{6}{7.2}{\rmdefault}{\mddefault}{\updefault}{\color[rgb]{0,0,.82}$\mathit{Node}(n^1_1, l^\epsilon_2, r^1_1)$}%
}}}}
\put(5063,-451){\makebox(0,0)[b]{\smash{{\SetFigFont{6}{7.2}{\rmdefault}{\mddefault}{\updefault}{\color[rgb]{0,0,0}$\mathit{NType}(r^\epsilon)$}%
}}}}
\put(3376,-2761){\makebox(0,0)[b]{\smash{{\SetFigFont{6}{7.2}{\rmdefault}{\mddefault}{\updefault}{\color[rgb]{0,0,0}$\mathit{NType}(n^0_2)$}%
}}}}
\put(6931,-2776){\makebox(0,0)[b]{\smash{{\SetFigFont{6}{7.2}{\rmdefault}{\mddefault}{\updefault}{\color[rgb]{0,0,0}$\mathit{NType}(n^1_1)$}%
}}}}
\put(10651,-1486){\makebox(0,0)[b]{\smash{{\SetFigFont{6}{7.2}{\rmdefault}{\mddefault}{\updefault}{\color[rgb]{0,0,.82}$\mathit{Node}(n^\epsilon_2, l^\epsilon_2, r^\epsilon_2)$}%
}}}}
\put(4501,-2686){\makebox(0,0)[b]{\smash{{\SetFigFont{6}{7.2}{\rmdefault}{\mddefault}{\updefault}{\color[rgb]{0,0,.82}$\mathit{Node}(n^0_2, l^0_2, r^\epsilon_1)$}%
}}}}
\put(1726,-2686){\makebox(0,0)[b]{\smash{{\SetFigFont{6}{7.2}{\rmdefault}{\mddefault}{\updefault}{\color[rgb]{0,0,.82}$\mathit{Node}(n^0_1, l^\epsilon_1, r^0_1)$}%
}}}}
\put(4201,-1486){\makebox(0,0)[b]{\smash{{\SetFigFont{6}{7.2}{\rmdefault}{\mddefault}{\updefault}{\color[rgb]{0,0,.82}$\mathit{Node}(n^\epsilon_1, l^\epsilon_1, r^\epsilon_1)$}%
}}}}
\put(11101,-361){\makebox(0,0)[b]{\smash{{\SetFigFont{6}{7.2}{\rmdefault}{\mddefault}{\updefault}{\color[rgb]{0,0,.82}$\mathit{Root}()$}%
}}}}
\put(376,-4111){\makebox(0,0)[b]{\smash{{\SetFigFont{5}{6.0}{\rmdefault}{\mddefault}{\updefault}{\color[rgb]{0,0,0}$\mathit{out}$}%
}}}}
\put(1951,-4111){\makebox(0,0)[b]{\smash{{\SetFigFont{5}{6.0}{\rmdefault}{\mddefault}{\updefault}{\color[rgb]{0,0,0}$\mathit{out}$}%
}}}}
\put(3526,-4111){\makebox(0,0)[b]{\smash{{\SetFigFont{5}{6.0}{\rmdefault}{\mddefault}{\updefault}{\color[rgb]{0,0,0}$\mathit{out}$}%
}}}}
\put(5101,-4111){\makebox(0,0)[b]{\smash{{\SetFigFont{5}{6.0}{\rmdefault}{\mddefault}{\updefault}{\color[rgb]{0,0,0}$\mathit{out}$}%
}}}}
\put(6751,-4111){\makebox(0,0)[b]{\smash{{\SetFigFont{5}{6.0}{\rmdefault}{\mddefault}{\updefault}{\color[rgb]{0,0,0}$\mathit{out}$}%
}}}}
\put(8326,-4111){\makebox(0,0)[b]{\smash{{\SetFigFont{5}{6.0}{\rmdefault}{\mddefault}{\updefault}{\color[rgb]{0,0,0}$\mathit{out}$}%
}}}}
\put(9901,-4111){\makebox(0,0)[b]{\smash{{\SetFigFont{5}{6.0}{\rmdefault}{\mddefault}{\updefault}{\color[rgb]{0,0,0}$\mathit{out}$}%
}}}}
\put(5176,-811){\makebox(0,0)[b]{\smash{{\SetFigFont{5}{6.0}{\rmdefault}{\mddefault}{\updefault}{\color[rgb]{0,0,0}$\mathit{req}$}%
}}}}
\put(1576,-1111){\makebox(0,0)[b]{\smash{{\SetFigFont{5}{6.0}{\rmdefault}{\mddefault}{\updefault}{\color[rgb]{0,0,0}$\mathit{reply}$}%
}}}}
\put(7876,-1111){\makebox(0,0)[b]{\smash{{\SetFigFont{5}{6.0}{\rmdefault}{\mddefault}{\updefault}{\color[rgb]{0,0,0}$\mathit{reply}$}%
}}}}
\put(2026,-1936){\makebox(0,0)[b]{\smash{{\SetFigFont{5}{6.0}{\rmdefault}{\mddefault}{\updefault}{\color[rgb]{0,0,0}$\mathit{req}$}%
}}}}
\put(8326,-1936){\makebox(0,0)[b]{\smash{{\SetFigFont{5}{6.0}{\rmdefault}{\mddefault}{\updefault}{\color[rgb]{0,0,0}$\mathit{req}$}%
}}}}
\put(901,-2386){\makebox(0,0)[b]{\smash{{\SetFigFont{5}{6.0}{\rmdefault}{\mddefault}{\updefault}{\color[rgb]{0,0,0}$\mathit{reply}$}%
}}}}
\put(3151,-2386){\makebox(0,0)[b]{\smash{{\SetFigFont{5}{6.0}{\rmdefault}{\mddefault}{\updefault}{\color[rgb]{0,0,0}$\mathit{reply}$}%
}}}}
\put(7201,-2386){\makebox(0,0)[b]{\smash{{\SetFigFont{5}{6.0}{\rmdefault}{\mddefault}{\updefault}{\color[rgb]{0,0,0}$\mathit{reply}$}%
}}}}
\put(9301,-2386){\makebox(0,0)[b]{\smash{{\SetFigFont{5}{6.0}{\rmdefault}{\mddefault}{\updefault}{\color[rgb]{0,0,0}$\mathit{reply}$}%
}}}}
\put(826,-3136){\makebox(0,0)[b]{\smash{{\SetFigFont{5}{6.0}{\rmdefault}{\mddefault}{\updefault}{\color[rgb]{0,0,0}$\mathit{req}$}%
}}}}
\put(3601,-3136){\makebox(0,0)[b]{\smash{{\SetFigFont{5}{6.0}{\rmdefault}{\mddefault}{\updefault}{\color[rgb]{0,0,0}$\mathit{req}$}%
}}}}
\put(7126,-3136){\makebox(0,0)[b]{\smash{{\SetFigFont{5}{6.0}{\rmdefault}{\mddefault}{\updefault}{\color[rgb]{0,0,0}$\mathit{req}$}%
}}}}
\put(9826,-3136){\makebox(0,0)[b]{\smash{{\SetFigFont{5}{6.0}{\rmdefault}{\mddefault}{\updefault}{\color[rgb]{0,0,0}$\mathit{req}$}%
}}}}
\put(  1,-3586){\makebox(0,0)[b]{\smash{{\SetFigFont{5}{6.0}{\rmdefault}{\mddefault}{\updefault}{\color[rgb]{0,0,0}$\mathit{reply}$}%
}}}}
\put(1126,-3586){\makebox(0,0)[b]{\smash{{\SetFigFont{5}{6.0}{\rmdefault}{\mddefault}{\updefault}{\color[rgb]{0,0,0}$\mathit{reply}$}%
}}}}
\put(3151,-3586){\makebox(0,0)[b]{\smash{{\SetFigFont{5}{6.0}{\rmdefault}{\mddefault}{\updefault}{\color[rgb]{0,0,0}$\mathit{reply}$}%
}}}}
\put(4276,-3586){\makebox(0,0)[b]{\smash{{\SetFigFont{5}{6.0}{\rmdefault}{\mddefault}{\updefault}{\color[rgb]{0,0,0}$\mathit{reply}$}%
}}}}
\put(6376,-3586){\makebox(0,0)[b]{\smash{{\SetFigFont{5}{6.0}{\rmdefault}{\mddefault}{\updefault}{\color[rgb]{0,0,0}$\mathit{reply}$}%
}}}}
\put(7501,-3586){\makebox(0,0)[b]{\smash{{\SetFigFont{5}{6.0}{\rmdefault}{\mddefault}{\updefault}{\color[rgb]{0,0,0}$\mathit{reply}$}%
}}}}
\put(9526,-3586){\makebox(0,0)[b]{\smash{{\SetFigFont{5}{6.0}{\rmdefault}{\mddefault}{\updefault}{\color[rgb]{0,0,0}$\mathit{reply}$}%
}}}}
\put(10726,-3586){\makebox(0,0)[b]{\smash{{\SetFigFont{5}{6.0}{\rmdefault}{\mddefault}{\updefault}{\color[rgb]{0,0,0}$\mathit{reply}$}%
}}}}
\put(1351,-4111){\makebox(0,0)[b]{\smash{{\SetFigFont{6}{7.2}{\rmdefault}{\mddefault}{\updefault}{\color[rgb]{0,0,0}$\mathit{LType}(r^0_1)$}%
}}}}
\put(2926,-4111){\makebox(0,0)[b]{\smash{{\SetFigFont{6}{7.2}{\rmdefault}{\mddefault}{\updefault}{\color[rgb]{0,0,0}$\mathit{LType}(l^0_2)$}%
}}}}
\put(4501,-4111){\makebox(0,0)[b]{\smash{{\SetFigFont{6}{7.2}{\rmdefault}{\mddefault}{\updefault}{\color[rgb]{0,0,0}$\mathit{LType}(r^\epsilon_1)$}%
}}}}
\put(6151,-4111){\makebox(0,0)[b]{\smash{{\SetFigFont{6}{7.2}{\rmdefault}{\mddefault}{\updefault}{\color[rgb]{0,0,0}$\mathit{LType}(l^\epsilon_2)$}%
}}}}
\put(7726,-4111){\makebox(0,0)[b]{\smash{{\SetFigFont{6}{7.2}{\rmdefault}{\mddefault}{\updefault}{\color[rgb]{0,0,0}$\mathit{LType}(r^1_1)$}%
}}}}
\put(9301,-4111){\makebox(0,0)[b]{\smash{{\SetFigFont{6}{7.2}{\rmdefault}{\mddefault}{\updefault}{\color[rgb]{0,0,0}$\mathit{LType}(l^1_2)$}%
}}}}
\put(10951,-4111){\makebox(0,0)[b]{\smash{{\SetFigFont{6}{7.2}{\rmdefault}{\mddefault}{\updefault}{\color[rgb]{0,0,0}$\mathit{LType}(r^\epsilon_2)$}%
}}}}
\put(1868,-1591){\makebox(0,0)[b]{\smash{{\SetFigFont{6}{7.2}{\rmdefault}{\mddefault}{\updefault}{\color[rgb]{0,0,0}$\mathit{NType}(n^\epsilon_1)$}%
}}}}
\put(-449,-436){\makebox(0,0)[b]{\smash{{\SetFigFont{6}{7.2}{\rmdefault}{\mddefault}{\updefault}{\color[rgb]{1,0,0}\ref{rule:tll-root}}%
}}}}
\put(-449,-1561){\makebox(0,0)[b]{\smash{{\SetFigFont{6}{7.2}{\rmdefault}{\mddefault}{\updefault}{\color[rgb]{1,0,0}\ref{rule:tll1}}%
}}}}
\put(8033,-1591){\makebox(0,0)[b]{\smash{{\SetFigFont{6}{7.2}{\rmdefault}{\mddefault}{\updefault}{\color[rgb]{0,0,0}$\mathit{NType}(n^\epsilon_2)$}%
}}}}
\put(5926,-1561){\makebox(0,0)[b]{\smash{{\SetFigFont{6}{7.2}{\rmdefault}{\mddefault}{\updefault}{\color[rgb]{1,0,0}\ref{rule:tll1}}%
}}}}
\put(2701,-2761){\makebox(0,0)[b]{\smash{{\SetFigFont{6}{7.2}{\rmdefault}{\mddefault}{\updefault}{\color[rgb]{1,0,0}\ref{rule:tll4}}%
}}}}
\put(631,-2791){\makebox(0,0)[b]{\smash{{\SetFigFont{6}{7.2}{\rmdefault}{\mddefault}{\updefault}{\color[rgb]{0,0,0}$\mathit{NType}(n^0_1)$}%
}}}}
\put(-449,-2761){\makebox(0,0)[b]{\smash{{\SetFigFont{6}{7.2}{\rmdefault}{\mddefault}{\updefault}{\color[rgb]{1,0,0}\ref{rule:tll4}}%
}}}}
\put(-74,-3811){\makebox(0,0)[b]{\smash{{\SetFigFont{6}{7.2}{\rmdefault}{\mddefault}{\updefault}{\color[rgb]{0,0,.82}$\mathit{Leaf}(l^\epsilon_1)$}%
}}}}
\put(1501,-3811){\makebox(0,0)[b]{\smash{{\SetFigFont{6}{7.2}{\rmdefault}{\mddefault}{\updefault}{\color[rgb]{0,0,.82}$\mathit{Leaf}(r^0_1)$}%
}}}}
\put(-599,-3811){\makebox(0,0)[b]{\smash{{\SetFigFont{6}{7.2}{\rmdefault}{\mddefault}{\updefault}{\color[rgb]{1,0,0}\ref{rule:tll5}}%
}}}}
\put(976,-3811){\makebox(0,0)[b]{\smash{{\SetFigFont{6}{7.2}{\rmdefault}{\mddefault}{\updefault}{\color[rgb]{1,0,0}\ref{rule:tll5}}%
}}}}
\put(2551,-3811){\makebox(0,0)[b]{\smash{{\SetFigFont{6}{7.2}{\rmdefault}{\mddefault}{\updefault}{\color[rgb]{1,0,0}\ref{rule:tll5}}%
}}}}
\put(3076,-3811){\makebox(0,0)[b]{\smash{{\SetFigFont{6}{7.2}{\rmdefault}{\mddefault}{\updefault}{\color[rgb]{0,0,.82}$\mathit{Leaf}(l^0_2)$}%
}}}}
\put(4126,-3811){\makebox(0,0)[b]{\smash{{\SetFigFont{6}{7.2}{\rmdefault}{\mddefault}{\updefault}{\color[rgb]{1,0,0}\ref{rule:tll5}}%
}}}}
\put(4651,-3811){\makebox(0,0)[b]{\smash{{\SetFigFont{6}{7.2}{\rmdefault}{\mddefault}{\updefault}{\color[rgb]{0,0,.82}$\mathit{Leaf}(r^\epsilon_1)$}%
}}}}
\put(5776,-3811){\makebox(0,0)[b]{\smash{{\SetFigFont{6}{7.2}{\rmdefault}{\mddefault}{\updefault}{\color[rgb]{1,0,0}\ref{rule:tll5}}%
}}}}
\put(7351,-3811){\makebox(0,0)[b]{\smash{{\SetFigFont{6}{7.2}{\rmdefault}{\mddefault}{\updefault}{\color[rgb]{1,0,0}\ref{rule:tll5}}%
}}}}
\put(8926,-3811){\makebox(0,0)[b]{\smash{{\SetFigFont{6}{7.2}{\rmdefault}{\mddefault}{\updefault}{\color[rgb]{1,0,0}\ref{rule:tll5}}%
}}}}
\put(10576,-3811){\makebox(0,0)[b]{\smash{{\SetFigFont{6}{7.2}{\rmdefault}{\mddefault}{\updefault}{\color[rgb]{1,0,0}\ref{rule:tll5}}%
}}}}
\put(5926,-2761){\makebox(0,0)[b]{\smash{{\SetFigFont{6}{7.2}{\rmdefault}{\mddefault}{\updefault}{\color[rgb]{1,0,0}\ref{rule:tll4}}%
}}}}
\put(9076,-2761){\makebox(0,0)[b]{\smash{{\SetFigFont{6}{7.2}{\rmdefault}{\mddefault}{\updefault}{\color[rgb]{1,0,0}\ref{rule:tll4}}%
}}}}
\put(6301,-3811){\makebox(0,0)[b]{\smash{{\SetFigFont{6}{7.2}{\rmdefault}{\mddefault}{\updefault}{\color[rgb]{0,0,.82}$\mathit{Leaf}(l^\epsilon_2)$}%
}}}}
\put(7876,-3811){\makebox(0,0)[b]{\smash{{\SetFigFont{6}{7.2}{\rmdefault}{\mddefault}{\updefault}{\color[rgb]{0,0,.82}$\mathit{Leaf}(r^1_1)$}%
}}}}
\put(9451,-3811){\makebox(0,0)[b]{\smash{{\SetFigFont{6}{7.2}{\rmdefault}{\mddefault}{\updefault}{\color[rgb]{0,0,.82}$\mathit{Leaf}(l^1_2)$}%
}}}}
\put(11101,-3811){\makebox(0,0)[b]{\smash{{\SetFigFont{6}{7.2}{\rmdefault}{\mddefault}{\updefault}{\color[rgb]{0,0,.82}$\mathit{Leaf}(r^\epsilon_2)$}%
}}}}
\put(11551,-4111){\makebox(0,0)[b]{\smash{{\SetFigFont{5}{6.0}{\rmdefault}{\mddefault}{\updefault}{\color[rgb]{0,0,0}$\mathit{out}$}%
}}}}
\put(10351,-4111){\makebox(0,0)[b]{\smash{{\SetFigFont{5}{6.0}{\rmdefault}{\mddefault}{\updefault}{\color[rgb]{0,0,0}$\mathit{in}$}%
}}}}
\put(8701,-4111){\makebox(0,0)[b]{\smash{{\SetFigFont{5}{6.0}{\rmdefault}{\mddefault}{\updefault}{\color[rgb]{0,0,0}$\mathit{in}$}%
}}}}
\put(7126,-4111){\makebox(0,0)[b]{\smash{{\SetFigFont{5}{6.0}{\rmdefault}{\mddefault}{\updefault}{\color[rgb]{0,0,0}$\mathit{in}$}%
}}}}
\put(5551,-4111){\makebox(0,0)[b]{\smash{{\SetFigFont{5}{6.0}{\rmdefault}{\mddefault}{\updefault}{\color[rgb]{0,0,0}$\mathit{in}$}%
}}}}
\put(3901,-4111){\makebox(0,0)[b]{\smash{{\SetFigFont{5}{6.0}{\rmdefault}{\mddefault}{\updefault}{\color[rgb]{0,0,0}$\mathit{in}$}%
}}}}
\put(2326,-4111){\makebox(0,0)[b]{\smash{{\SetFigFont{5}{6.0}{\rmdefault}{\mddefault}{\updefault}{\color[rgb]{0,0,0}$\mathit{in}$}%
}}}}
\put(751,-4111){\makebox(0,0)[b]{\smash{{\SetFigFont{5}{6.0}{\rmdefault}{\mddefault}{\updefault}{\color[rgb]{0,0,0}$\mathit{in}$}%
}}}}
\put(-224,-4111){\makebox(0,0)[b]{\smash{{\SetFigFont{6}{7.2}{\rmdefault}{\mddefault}{\updefault}{\color[rgb]{0,0,0}$\mathit{LType}(l^\epsilon_1)$}%
}}}}
\put(-749,-4111){\makebox(0,0)[b]{\smash{{\SetFigFont{5}{6.0}{\rmdefault}{\mddefault}{\updefault}{\color[rgb]{0,0,0}$\mathit{in}$}%
}}}}
\put(10801,-2686){\makebox(0,0)[b]{\smash{{\SetFigFont{6}{7.2}{\rmdefault}{\mddefault}{\updefault}{\color[rgb]{0,0,.82}$\mathit{Node}(n^1_2, l^1_2, r^\epsilon_2)$}%
}}}}
\put(9631,-2776){\makebox(0,0)[b]{\smash{{\SetFigFont{6}{7.2}{\rmdefault}{\mddefault}{\updefault}{\color[rgb]{0,0,0}$\mathit{NType}(n^1_2)$}%
}}}}
\end{picture}%

%% file: tll-auto.pdf_t
\begin{picture}(0,0)%
\includegraphics{tll-auto.pdf}%
\end{picture}%
\setlength{\unitlength}{1973sp}%
\begingroup\makeatletter\ifx\SetFigFont\undefined%
\gdef\SetFigFont#1#2#3#4#5{%
  \reset@font\fontsize{#1}{#2pt}%
  \fontfamily{#3}\fontseries{#4}\fontshape{#5}%
  \selectfont}%
\fi\endgroup%
\begin{picture}(9609,2802)(604,-2173)
\put(1201,-136){\makebox(0,0)[b]{\smash{{\SetFigFont{6}{7.2}{\rmdefault}{\mddefault}{\updefault}{\color[rgb]{0,0,0}$q^\downarrow_{\text{\ref{rule:tll-root}},l_1}$}%
}}}}
\put(5401,-1594){\makebox(0,0)[b]{\smash{{\SetFigFont{6}{7.2}{\rmdefault}{\mddefault}{\updefault}{\color[rgb]{0,0,0}$q^\downarrow_{\text{\ref{rule:tll5}},n}$}%
}}}}
\put(7531,-1624){\makebox(0,0)[b]{\smash{{\SetFigFont{6}{7.2}{\rmdefault}{\mddefault}{\updefault}{\color[rgb]{0,0,0}$q^\downarrow_{\text{\ref{rule:tll4}},r}$}%
}}}}
\put(3331,-154){\makebox(0,0)[b]{\smash{{\SetFigFont{6}{7.2}{\rmdefault}{\mddefault}{\updefault}{\color[rgb]{0,0,0}$q^\downarrow_{\text{\ref{rule:tll1}},l}$}%
}}}}
\put(3346,-1609){\makebox(0,0)[b]{\smash{{\SetFigFont{6}{7.2}{\rmdefault}{\mddefault}{\updefault}{\color[rgb]{0,0,0}$q^\downarrow_{\text{\ref{rule:tll4}},l}$}%
}}}}
\put(7516,-169){\makebox(0,0)[b]{\smash{{\SetFigFont{6}{7.2}{\rmdefault}{\mddefault}{\updefault}{\color[rgb]{0,0,0}$q^\downarrow_{\text{\ref{rule:tll1}},r}$}%
}}}}
\put(1426,-811){\makebox(0,0)[b]{\smash{{\SetFigFont{6}{7.2}{\rmdefault}{\mddefault}{\updefault}{\color[rgb]{0,0,0}$(1,\downarrow)$}%
}}}}
\put(3001,-811){\makebox(0,0)[b]{\smash{{\SetFigFont{6}{7.2}{\rmdefault}{\mddefault}{\updefault}{\color[rgb]{0,0,0}$(1,\downarrow)$}%
}}}}
\put(4351,-1411){\makebox(0,0)[b]{\smash{{\SetFigFont{6}{7.2}{\rmdefault}{\mddefault}{\updefault}{\color[rgb]{0,0,0}$(1,\downarrow)$}%
}}}}
\put(7201,-811){\makebox(0,0)[b]{\smash{{\SetFigFont{6}{7.2}{\rmdefault}{\mddefault}{\updefault}{\color[rgb]{0,0,0}$(2,\downarrow)$}%
}}}}
\put(6451,-1411){\makebox(0,0)[b]{\smash{{\SetFigFont{6}{7.2}{\rmdefault}{\mddefault}{\updefault}{\color[rgb]{0,0,0}$(2,\downarrow)$}%
}}}}
\put(3526,389){\makebox(0,0)[b]{\smash{{\SetFigFont{6}{7.2}{\rmdefault}{\mddefault}{\updefault}{\color[rgb]{0,0,0}$(1,\downarrow)$}%
}}}}
\put(7276,389){\makebox(0,0)[b]{\smash{{\SetFigFont{6}{7.2}{\rmdefault}{\mddefault}{\updefault}{\color[rgb]{0,0,0}$(2,\downarrow)$}%
}}}}
\put(8551, 14){\makebox(0,0)[b]{\smash{{\SetFigFont{6}{7.2}{\rmdefault}{\mddefault}{\updefault}{\color[rgb]{0,0,0}$(2,\downarrow)$}%
}}}}
\put(2251, 14){\makebox(0,0)[b]{\smash{{\SetFigFont{6}{7.2}{\rmdefault}{\mddefault}{\updefault}{\color[rgb]{0,0,0}$(1,\downarrow)$}%
}}}}
\put(9151,-811){\makebox(0,0)[b]{\smash{{\SetFigFont{6}{7.2}{\rmdefault}{\mddefault}{\updefault}{\color[rgb]{0,0,0}$(2,\downarrow)$}%
}}}}
\put(9616,-169){\makebox(0,0)[b]{\smash{{\SetFigFont{6}{7.2}{\rmdefault}{\mddefault}{\updefault}{\color[rgb]{0,0,0}$q^\downarrow_{\text{\ref{rule:tll-root}},r_1}$}%
}}}}
\end{picture}%

%% file: experiments.tex
\begin{tabular}{|l|c|c|c|c|c|c|c|c|}
\hline
benchmark & \#states/comp. & \#rules & \#inter. & deadlock & time (sec) & trap-inv & $1$-inv & $\kappa$ \\
\hline 
  {\sf ring} & $2 \times 2$ & 3 & 3 & \checkmark & 0.01 & \checkmark & - & 1 \\
  {\sf star} & $2 \times 2$ & 3 & 4 & \checkmark & 0.01 & \checkmark & - & 1 \\
  {\sf star-ring} & $2 \times 3 \times 3$ & 3 & 9 & \checkmark & 0.03 & \checkmark & - & 1 \\
  {\sf alt-philo-sym} & $3 \times 2$ & 3 & 9 & $\times$ & 0.70 & \checkmark & \checkmark & 1  \\
  {\sf alt-philo-asym} & $3 \times 2$ & 3 & 9 & \checkmark & 0.67 & \checkmark & \checkmark & 1 \\
  {\sf sync-philo} & $2 \times 2$ & 3 & 6 & \checkmark & 0.03 & \checkmark & - & 1 \\
  {\sf tree-dfs} & $2 \times 6 \times 2$ & 4 & 6 & \checkmark & 0.07 & \checkmark & - & 2 \\
  {\sf tree-back-root} & $2 \times 2$ & 3 & 5 & \checkmark & 0.03 & \checkmark & - & 2 \\
  {\sf tree-linked-leaves} & $2 \times 2 \times 4 \times 3$ & 4  & 10 & \checkmark & 0.27 & \checkmark & - & 2 \\
\hline 
\end{tabular}